\input harvmac
\input epsf
\noblackbox
\def\npb#1#2#3{{\it Nucl.\ Phys.} {\bf B#1} (#2) #3}
\def\plb#1#2#3{{\it Phys.\ Lett.} {\bf B#1} (#2) #3}
\def\prl#1#2#3{{\it Phys.\ Rev.\ Lett.} {\bf #1} (#2) #3}
\def\prd#1#2#3{{\it Phys.\ Rev.} {\bf D#1} (#2) #3}
\def\rmp#1#2#3{{\it Rev.\ Mod.\ Phys.} {\bf #1} (#2) #3}
\def\mpla#1#2#3{{\it Mod.\ Phys.\ Lett.} {\bf A#1} (#2) #3}

\def\jmp#1#2#3{{\it J. Math.\ Phys.} {\bf #1} (#2) #3}

\def\aop#1#2#3{{\it Ann.\ of Phys.} {\bf #1} (#2) #3}
\def\jhep#1#2#3{{\it JHEP\/} {\bf #1} (#2) #3}
\def\atmp#1#2#3{{\it Adv.\ Theor.\ Math.\ Phys.} {\bf #1} (#2) #3}
\newcount\figno
\figno=0
\def\fig#1#2#3{
\par\begingroup\parindent=0pt\leftskip=1cm\rightskip=1cm\parindent=0pt
\baselineskip=11pt
\global\advance\figno by 1
\midinsert
\epsfxsize=#3
\centerline{\epsfbox{#2}}
\vskip 12pt
{\bf Fig.\ \the\figno: } #1\par
\endinsert\endgroup\par
}
\def\figlabel#1{\xdef#1{\the\figno}}
\def\encadremath#1{\vbox{\hrule\hbox{\vrule\kern8pt\vbox{\kern8pt
\hbox{$\displaystyle #1$}\kern8pt}
\kern8pt\vrule}\hrule}}
\def\bberkeley{\centerline{\it Berkeley Center for Theoretical Physics and 
Department of Physics}
\centerline{\it University of California, Berkeley, CA 94720-7300}
\centerline{\it and}
\centerline{\it Theoretical Physics Group, Lawrence Berkeley National 
Laboratory}
\centerline{\it Berkeley, CA 94720-8162, USA}}

\def\frac#1#2{{#1 \over #2}}

\def\p{\partial}
\def\semi{\subset\kern-1em\times\;}

                   \def\CG{{\cal G}}
\def\CH{{\cal H}}                   
                   
                   \def\CN{{\cal N}}
\def\CO{{\cal O}}                   \def\CP{{\cal P}}
\def\CR{{\cal R}}                   \def\CT{{\cal T}}
                   
\def\CW{{\cal W}}                   \def\CZ{{\cal Z}}
                     \def\Z{{\bf Z}}
\def\sla{{\sl a}}
\def\slb{{\sl b}}
\def\Re{{\,{\rm Re}\,}}
\def\Im{{\,{\rm Im}\,}}
\def\Li{{\,\rm Li}}
\def\Res{{\,\rm Res}}
\def\ie{{\it i.e.}}
\def\eg{{\it e.g.}}

\def\subsubsec#1{\bigskip\noindent
{\underbar{\it #1}}\medskip}
\def\hepth#1{[arXiv:hep-th/{#1}]}
\def\condmat#1{[arXiv:cond-mat/{#1}]}
\Title{\vbox{\baselineskip12pt
\hbox{hep-th/0508024}}}
{\vbox{\centerline{Noncritical M-Theory in $2+1$ Dimensions}
\bigskip
\centerline{as a Nonrelativistic Fermi Liquid}}}
\bigskip
\font\authfont=cmr12
\centerline{\authfont Petr Ho\v rava and Cynthia A. Keeler}
\medskip\bigskip\medskip
\baselineskip14pt
\bberkeley
\medskip\bigskip\medskip\bigskip
\centerline{\bf Abstract}
\bigskip
\noindent
We claim that the dynamics of noncritical string theories in two dimensions 
is related to an underlying noncritical version of M-theory, which we define 
in terms of a double-scaled nonrelativistic Fermi liquid in $2+1$ dimensions.  
After reproducing Type 0A and 0B string theories as solutions, we study the 
natural M-theory vacuum.   The vacuum energy of this solution can be 
evaluated exactly, its form suggesting a duality to the Debye model of 
phonons in a melting solid, and a possible topological nature of the theory.  
The physical spacetime is emergent in this theory, only for states that admit 
a hydrodynamic description.  Among the solutions of the hydrodynamic equations 
of motion for the Fermi surface, we find families describing the decay of one 
two-dimensional string theory into another via an intermediate M-theory 
phase.  
\Date{July 2005}
\nref\newhat{M.R.~Douglas, I.R.~Klebanov, D.~Kutasov, J.~Maldacena and 
E.~Martinec, ``A New Hat for the $c=1$ Matrix Model'' \hepth{0307195}.}
\nref\tato{T. Takayanagi and N. Toumbas, ``A Matrix Model Dual of Type 0B 
String Theory in Two Dimensions,'' \jhep{0307}{2003}{064} \hepth{0307083}.}
\nref\gmrev{P. Ginsparg and G. Moore, ``Lectures on 2D Gravity and 2D String 
Theory'' \hepth{9304011}.}
\nref\igrev{I. Klebanov, ``String Theory in Two Dimensions'' \hepth{9108019}.}
\nref\polrev{J. Polchinski, ``What is String Theory?'' \hepth{9411028}.}
\nref\alexrev{S. Alexandrov, ``Matrix Quantum Mechanics and Two-dimensional 
String Theory in Non-trivial Backgrounds'' \hepth{0311273}.}
\nref\nakarev{Y. Nakayama, ``Liouville Field Theory -- A Decade after the 
Revolution'' \hepth{0402009}.}
\nref\emilrev{E.J. Martinec, ``The Annular Report on Non-Critical String 
Theory'' \hepth{0305148}, ``Matrix Models and 2D String Theory'' 
\hepth{0410136}.}
\nref\mmre{J. McGreevy and H. Verlinde, ``Strings from Tachyons: The $c=1$ 
Matrix Reloaded,'' \jhep{0312}{2003}{054} \hepth{0304224}.}
\nref\kazmig{V.A. Kazakov and A.A. Migdal, ``Recent Progress in the Theory of 
Noncritical Strings,'' \npb{311}{1988}{171}.}
\nref\grossmilj{D.J. Gross and N. Miljkovi\'c, ``A Nonperturbative Solution 
of $D=1$ String Theory,'' \plb{238}{1990}{217}.}
\nref\grosskl{D.J. Gross and I.R. Klebanov, ``One-Dimensional String Theory 
on a Circle,'' \npb{344}{1990}{475}.}
\nref\ceeone{E. Br\'ezin, V.A. Kazakov and Al.B. Zamolodchikov, 
``Scaling Violation in a Field Theory of Closed Strings in One Physical 
Dimension,'' \npb{338}{1990}{673}\hfill\break
P. Ginsparg and J. Zinn-Justin, ``2-D Gravity + 1-D Matter,'' 
\plb{240}{1990}{333}.}
\nref\kazakov{V. Kazakov, ``Bosonic Strings and String Field Theories in 
One-Dimensional Target Space,'' LPTENS-90-30, in: {\it Random Surfaces and 
Quantum Gravity\/}, Carg\`ese 1990 Proceedings.}
\nref\moore{G. Moore, ``Double-Scaled Field Theory at $c=1$,'' 
\npb{368}{1992}{557}.}
\nref\deformed{A. Jevicki and T. Yoneya, ``A Deformed Matrix Model and the 
Black Hole Background in Two-Dimensional String Theory,'' \npb{411}{1994}{64} 
\hepth{9305109}.}
\nref\danielsson{U.H. Danielsson, ``A Matrix Model Black Hole,'' 
\npb{410}{1993}{395} \hepth{9306063}; 
``The Deformed Matrix Model at Finite Radius and a New Duality Symmetry,'' 
\plb{325}{1994}{33} \hepth{9309157}; 
``Two-Dimensional String Theory, Topological Field Theories and the 
Feformed Matrix Model,'' \npb{425}{1994}{261} \hepth{9401135}; 
``The Scattering of Strings in a Black-Hole Background,'' \plb{338}{1994}{158} 
\hepth{9405052}; 
``A Matrix Model Black Hole: Act II,'' \jhep{0402}{2004}{067} \hepth{0312203}; 
U.H. Danielsson, N. Johansson, M. Larfors, M.E. Olsson and M. Vonk, 
``4D Black Holes and Holomorphic Factorization of the 0A Matrix Model,'' 
\hepth{0506219}.}
\nref\demetrod{K. Demeterfi and J.P. Rodrigues, ``States and Quantum Effects 
in the Collective Field Theory of a Deformed Matrix Model'' \hepth{9306141}.}
\nref\demklerod{K. Demeterfi, I.R. Klebanov and J.P. Rodrigues, ``The Exact 
$S$-Matrix of the Deformed $c=1$ Matrix Model'' \hepth{9308036}.}
\nref\gtato{S. Gukov, T. Takayanagi and N. Toumbas, ``Flux Backgrounds in 
2D String Theory,'' \jhep{0403}{2004}{017} \hepth{0312208}.}
\nref\takaiia{T. Takayanagi, ``Comments on 2D Type IIA String and Matrix 
Model,'' \jhep{0411}{2004}{030} \hepth{0408086}.}
\nref\seibergiia{N. Seiberg, ``Observations on the Moduli Space of Two 
Dimensional String Theory,'' \jhep{0503}{2005}{010} \hepth{0502156}.}
\nref\oren{O. Bergman and M.R. Gaberdiel, ``Dualities of Type 0 Strings,'' 
\jhep{9907}{1999}{022} \hepth{9906055}.}
\nref\kfermi{P. Ho\v rava, ``Stability of Fermi Surfaces and K-Theory,'' 
\prl{95}{2005}{016405}, \hepth{0503006}.}
\nref\yaffe{L.G. Yaffe, ``Large $N$ Limits as Classical Mechanics,'' 
\rmp{54}{1982}{407}.}
\nref\hw{P. Ho\v rava and E. Witten, ``Heterotic and Type I String Dynamics 
from Eleven Dimensions,'' \npb{460}{1996}{506} \hepth{9510209}, 
``Eleven-Dimensional Supergravity on a 
Manifold with Boundary,'' \npb{475}{1996}{94} \hepth{9603142}.}
\nref\highnon{D. Kutasov and N. Seiberg, ``Noncritical Superstrings,'' 
\plb{251}{1990}{67}\hfill\break
S. Murthy, ``Notes on Noncritical Superstrings in Various Dimensions,'' 
\jhep{0311}{2003}{056} \hepth{0305197}.}
\nref\hft{P. Ho\v rava, ``M-Theory as a Holographic Field Theory,'' 
\prd{59}{1999}{046004} \hepth{9712130}.}
\nref\phdm{P. Ho\v rava and D. Minic, ``Probable Values of the Cosmological 
Constant in a Holographic Theory,'' \prl{85}{2000}{1610} \hepth{0001145}.}
\nref\debye{P. Debye, ``Zur Theorie der Spezifischen W\"arme,'' {\it Annalen 
der Physik\/} {\bf 39} (1912) 789.}
\nref\lebellac{see, \eg , \S{5} of M. Le~Bellac, F. Mortessagne and 
G.G. Batrouni, {\it Equilibrium and Non-Equilibrium Statistical 
Thermodynamics} (CUP, Cambridge, 2004).}
\nref\melting{A. Okounkov, N. Reshetikhin and C. Vafa, ``Quantum Calabi-Yau 
and Classical Crystals'' \hepth{0309208}.}
\nref\djcoll{S.R. Das and A. Jevicki, ``String Field Theory and Physical 
Interpretation of $D=1$ Strings,'' \mpla{5}{1990}{1639}.}
\nref\haldane{F.D.M. Haldane, in: {\it Perspectives in Many-Particle 
Physics}, eds: R.A. Broglia and J.R. Schrieffer (North Holland, 1994).}
\nref\bosonization{A. Houghton and J.B. Marston, ``Bosonization and Fermion 
Liquids in Dimensions Greater Than One,'' \condmat{9210007}; 
A. Houghton, H.-J. Kwon and J.B. Marston, 
``Multidimensional Bosonization,'' \condmat{9810388}\hfill\break
A.H.~Castro~Neto and E.~Fradkin, ``Bosonization of the Low Energy 
Excitations of Fermi Liquids'' \condmat{9304014},
``Bosonization of Fermi Liquids'' \hfill\break  \condmat{9307005},  
``Exact Solution of the Landau Fixed Point via Bosonization'' 
\condmat{9310046}.}
\nref\shankar{R. Shankar, ``Renormalization Group Approach to Interacting 
Fermions,'' \rmp{66}{1994}{129}.}
\nref\moorepr{G. Moore, M.R. Plesser and S. Ramgoolam, ``Exact S-Matrix for 
2D String Theory,'' \npb{377}{1992}{143} \hepth{9111035}.}
\nref\szerob{O. DeWolfe, R. Roiban, M. Spradlin, A. Volovich and J. Walcher, 
``On the $S$-Matrix of Type 0 String Theory,'' \jhep{0311}{2003}{012} 
\hepth{0309148}.}
\nref\grring{E. Witten, ``Ground Ring of Two Dimensional String Theory,'' 
\npb{373}{1992}{187} \hepth{9108004}\hfill\break
E. Witten and B. Zwiebach, ``Algebraic Structures and Differential 
Geometry in 2D String Theory,'' \npb{377}{1992}{55}.}
\nref\polceeone{J. Polchinski, ``Critical Behavior of Random Surfaces in 
One Dimension,'' \npb{346}{1990}{253}; ``Classical Limit of 
$(1+1)$-Dimensional String Theory,'' \npb{362}{1991}{125}.}
\nref\djordje{D. Minic, J. Polchinski and Z. Yang, ``Translation Invariant 
Backgrounds in $(1+1)$-Dimensional String Theory,'' \npb{369}{1992}{324}.}
\nref\kapustin{A. Kapustin, ``Noncritical Superstrings in a Ramond-Ramond 
Background,'' \jhep{0406}{2004}{024} \hepth{0308119}.}
\nref\hermanads{H. Verlinde, ``Superstrings on $AdS_2$ and Superconformal 
Matrix Quantum Mechanics'' \hepth{0403024}.}
\nref\andyads{A. Strominger, ``A Matrix Model for $AdS_2$,'' 
\jhep{0403}{2004}{066} \hepth{0312194}.}
\nref\schw{J. Schwinger, ``Brownian Motion of a Quantum Oscillator,'' 
\jmp{2}{1961}{407}\hfill\break
L.V. Keldysh, {\it Sov.\ Phys. JETP\/} {\bf 20} (1964) 1018.}
\nref\niemisem{A.J. Niemi and G.W. Semenoff, ``Finite Temperature Quantum 
Field Theory in Minkowski Space,'' \aop{152}{1984}{105}, ``Thermodynamic 
Calculations in Relativistic Finite-Temperature Quantum Field Theories,'' 
\npb{230[FS10]}{1984}{181}.}
\nref\tfd{H. Umezawa, H. Matsumoto and M. Tachiki, {\it Thermo Field Dynamics 
and Condensed States\/} (North Holland, Amsterdam, 1982).}
\nref\maldath{J.M. Maldacena, ``Eternal Black Holes in AdS,'' 
\jhep{0304}{2003}{021}, \hepth{0106112}.}
\nref\veronika{L. Fidkowski, V. Hubeny, M. Kleban and S. Shenker, ``The Black 
Hole Singularity in AdS/CFT,'' \jhep{0402}{2004}{014}, \hepth{0306170}.}
\nref\itzmcg{N. Itzhaki and J. McGreevy, ``The Large $N$ Harmonic Oscillator 
as a String Theory,''\prd{71}{2005}{025003} \hepth{0408180}.}
\nref\boyar{A. Boyarsky, V.V. Cheianov and O. Ruchayskiy, ``Fermions in 
the Harmonic Potential and String Theory,'' \jhep{0501}{2005}{010} 
\hepth{0409129}.}
\nref\cjr{S. Corley, A. Jevicki and S. Ramgoolam, ``Exact Correlators of 
Giant Gravitons from Dual $N=4$ SYM,'' \atmp{5}{2002}{809} \hepth{0111222}.}
\nref\berenstein{D. Berenstein, ``A Toy Model for the AdS/CFT 
Correspondence,'' \jhep{0407}{2004}{018} \hepth{0403110}.}
\nref\juannati{J. Maldacena and N. Seiberg, ``Flux-vacua in Two Dimensional 
String Theory'' \hepth{0506141}.}
\nref\karczs{J.L. Karczmarek and A. Strominger, ``Matrix Cosmology'' 
\jhep{0404}{2004}{055} \hepth{0309138}, ``Closed String Tachyon Condensation 
at $c=1$,'' \jhep{0405}{2004}{062} \hepth{0403169}.}
\nref\sumit{S.R. Das, J.L. Davis, F. Larsen and P. Mukhopadhyay, ``Particle 
Production in Matrix Cosmology,'' \prd{70}{2004}{044017} \hepth{0403275}.}
\nref\karczms{J.L. Karczmarek, A. Maloney and A. Strominger, ``Hartle-Hawking 
Vacuum for $c=1$ Tachyon Condensation,'' \jhep{0412}{2004}{027} 
\hepth{0405092}.}
\nref\daskarcz{S.R. Das and J.L. Karczmarek, ``Spacelike Boundaries from the 
$c=1$ Matrix Model,'' \prd{71}{2005}{086006} \hepth{0412093}.}
\nref\das{S.R. Das, ``D Branes in 2d String Theory and Classical Limits'' 
\hepth{0401067}, ``Non-trivial 2d Space-times from Matrices'' 
\hepth{0503002}.}
\nref\bfss{T. Banks, W. Fischler, S.H. Shenker and L. Susskind, ``M-Theory as 
a Matrix Model: A Conjecture'' \prd{55}{1997}{5112} \hepth{9610043}.}
\nref\ssscale{A. Sen, ``D0-Branes on $T^n$ and Matrix Theory,'' 
\atmp{2}{1998}{51} \hepth{9709220}\hfill\break
N.Seiberg, ``Why is the Matrix Model Correct?'' \prl{79}{1997}{3577} 
\hepth{9710009}.}
\nref\polmm{J. Polchinski, ``M-Theory and the Light Cone'' \hepth{9903165}.}
\nref\phmx{P. Ho\v rava, ``Type IIA D-Branes, K-Theory, and Matrix Theory,'' 
\atmp{2}{1999}{1373} \hepth{9812135}.}
\nref\minaetal{M. Aganagic, R. Dijkgraaf, A. Klemm, M. Mari\~ no and C. Vafa, 
``Topological Strings and Integrable Hierarchies'' \hepth{0312085}.}
\nref\topom{R. Dijkgraaf, S. Gukov, A. Neitzke and C. Vafa, ``Topological 
M-Theory as Unification of Form Theories of Gravity'' \hepth{0411073}.}
\nref\ewcs{E. Witten, ``2+1 Dimensional Gravity as an Exactly Soluble 
System,'' \npb{311}{1988}{46}; ``Topology-Changing Amplitudes in $2+1$ 
Dimensional Gravity,'' \npb{323}{1989}{113}.}
\nref\carlip{S. Carlip, {\it Quantum Gravity in 2\/+\/1 Dimensions} 
(CUP, Cambridge, 1998).}
\nref\mach{E. Mach, {\it Die Mechanik in ihrer Entwicklung} (Brockhaus, 
Leipzig, 1883).}
\newsec{Introduction and Summary}

In the wake of the second string revolution ten years ago, we have been  
left with a satisfying picture of a unique theory, with different string 
vacua connected by a web of dualities. It is somewhat ironic, however, that 
in the process of establishing that string theory is a unique theory, it was 
also discovered that this unique theory -- provisionally called ``M-theory'' 
-- is not always a theory of fundamental strings.  Despite much progress in 
our understanding of M-theory in the last ten years, the nature of its degrees 
of freedom is still rather elusive, representing one of the major challenges 
of the field.  

Ultimately, we wish to understand the landscape of all possible 
solutions of the theory.  However, it is difficult to imagine how this would 
be possible in the absence of a clear understanding of the nature of the 
underlying degrees of freedom.  
On another note, it has long been suspected that the physical spacetime in 
quantum gravity should emerge as a derived concept.  A more precise 
realization of this hope would also seem to require access to more fundamental 
degrees of freedom of quantum gravity.  

In this paper, we will address these issues in the highly controlled (indeed, 
exactly solvable) context of noncritical string theories in two spacetime 
dimensions \refs{\newhat-\emilrev}, as defined via their matrix model 
formulation \refs{\mmre-\ceeone}. We shall find that 
noncritical string theories are also connected in a larger framework, of a 
theory in $2+1$ dimensions which we refer to as ``noncritical M-theory''.  
We give an exact, nonperturbative definition of this noncritical M-theory, 
from which many exact results can be obtained.  In the process, we will get 
our first glimpse into the fundamental degrees of freedom in M-theory, at 
least in its $2+1$-dimensional incarnation: Noncritical M-theory is a theory 
of double-scaled nonrelativistic fermions in $2+1$ dimensions.  This exact 
formulation of noncritical M-theory will allow us to understand in detail 
the entire space of solutions of the theory, the space frequently represented 
in full M-theory by the well-known ``starfish'' diagram.  

The organization and outline of this paper are as follows.  After a 
brief review of noncritical Type 0A and 0B strings in $1+1$ dimensions in 
Section~2.1, we present our definition of noncritical M-theory in terms of 
a double-scaled Fermi liquid in $2+1$ dimensions in Section~2.2.  In 
particular, we propose to identify the extra dimension of M-theory with 
the angular dimension on the plane populated by the nonrelativistic 
fermions.  The theory is further developed in Section~3, where we also discuss 
the moduli space of all solutions of the theory, as well as the connection 
between the existence of hydrodynamic degrees of freedom and the existence 
of a semiclassical spacetime description of a given solution.  In Section~4, 
we reproduce the linear dilaton vacua of two-dimensional Type 0A and 0B 
noncritical string theories as solutions of noncritical M-theory.  In 
Section~5, we introduce the natural M-theory vacuum.  First we analyze the 
scaling at the leading order in large $N$ and identify the natural scaling 
variable $\mu$, and then define the nonperturbative 
double-scaling limit of this vacuum.  

Section~6 contains some of the central results of this paper.  In particular, 
we present an exact calculation of the vacuum energy of the M-theory vacuum 
solution, as a function of the scaling variable $\mu$. The exact formula for 
the vacuum energy turns out to be one-loop exact (in perturbation theory in 
the powers of $1/\mu\sim\kappa^{2/3}$), with an infinite series of 
instanton-like corrections, each of which is also one-loop exact.  This result 
is suggestive of a possible topological nature, or at least localization of 
the path integral, of noncritical M-theory.  In Section~7 we point out that 
the exact formula for the vacuum energy suggests a dual interpretation, 
in terms of the Debye model of a quantum crystal at finite temperature set by 
the string scale.  In fact, $\mu$ controls how many atoms have been {\it 
removed\/} from a large Debye crystal, leading to an interpretation in terms 
of crystal melting.  

In Section~8 we address two more general aspects of noncritical M-theory: 
Its observables and symmetries.  A particularly natural observable is given 
by the density of eigenvalues.  This observable is the M-theory analog of the 
massless tachyon from noncritical string theories.  The theory is shown 
to exhibit an infinite $\CW$ symmetry algebra.  Section~9 develops a general 
framework for identifying ``good'' hydrodynamic solutions of the theory, 
for which a spacetime description should be possible.  We formulate the 
classical hydrodynamic equation of motion for the Fermi surface, and present 
several simple static solutions of this equation.  A surprising duality to 
the thermofield dynamics of fermions in the rightside-up harmonic oscillator 
potential is found.  Section~10 continues the analysis by introducing a 
general class of time-dependent solutions of the Fermi surface equations of 
motion.  Among the time-dependent solutions, we find classes representing 
a dynamical change of the spacetime dimension.  In particular, there are 
solutions describing the decay of a $1+1$-dimensional string theory vacuum 
to another one via an intermediate $2+1$-dimensional M-theory phase.  
Section~11 concludes with some general remarks and some open questions.

\newsec{From Noncritical Strings to Noncritical M-Theory}

Our starting point is the matrix model formulation of various noncritical 
strings in two spacetime dimensions.  We concentrate on the Type 0A and 
0B superstrings \refs{\newhat,\tato}, but our analysis can be easily extended 
to include other vacua, such as the Type II or bosonic strings in two 
dimensions.  

\subsec{Type 0A and 0B Strings in Two Dimensions}

This is not the right place for a lengthy overview of two-dimensional 
strings, and we only highlight some basic aspects as needed for the rest of 
the paper.  Excellent extensive reviews of the subject exist, see, \eg ,  
\refs{\gmrev-\emilrev}.  

Type 0 superstrings are defined via a double-scaling limit of the Euclidean 
matrix path integral 
\eqn\eepathmatrix{\CZ=\int DM(t)e^{-S(M)}.}
In Type 0B theory \refs{\tato,\newhat}, the action is given by
\eqn\eematsb{S_{0B}(M)=\beta N\int dt\,\Tr\,\left(\frac{1}{2}(D_tM)^2+V(M)
\right).}
$M$ is a Hermitian $N\times N$ matrix, $D_t$ is 
the covariant derivative with respect to a $U(N)$ gauge field $A_0$, and 
$\beta$ is a coupling constant which can be conveniently reabsorbed into 
$M$.  

The Type 0A superstring similarly corresponds to a quiver matrix mechanics 
\newhat , 
\eqn\eematsa{S_{0A}(M,M^\dagger)=\beta N\int dt\left(\Tr\,[(D_tM)^\dagger D_tM+
V(M,M^\dagger)]\right).}
In this case, $M$ is an $N\times (N+q)$ complex matrix, and the gauge group 
is $U(N)\times U(N+q)$.  $q$ is interpreted as the net D0-brane charge or, 
alternatively, the value of the RR two-form flux in the vacuum.  $M$ is the 
matrix of open-string tachyon modes on the system of $N+q$ D0-branes 
and $N$ anti D0-branes in Type 0A theory, or $N$ unstable D0-branes in 
the Type 0B matrix model, along the lines of \mmre .  

The universal part of the potential is 
\eqn\eeunivpot{V(M)=-\frac{1}{2}\omega_0^2M^2+\ldots.}
Here the ``$\ldots$'' stand for stabilizing, nonuniversal terms in the 
potential, and $\omega_0$ is the fundamental frequency scale of the theory.  
In Type 0A and 0B string theories, this fundamental frequency sets the string 
scale, $\omega_0=1/\sqrt{2\alpha'}$.  

In the singlet sector, the matrix models reduce to a theory of $N$ free 
fermions, representing the locations of $N$ eigenvalues 
$y_\alpha$, $\alpha=1,\ldots,\ N$ of $M$ along a spatial dimension $y$.  
The ground state of this system corresponds to all states filled up 
to a (negative) Fermi energy $\varepsilon_F$.  The second-quantized 
Hamiltonian is 
\eqn\eematham{\CH=\beta N\int dy\left(-\frac{1}{2(\beta N)^2}
\p_y\psi^\dagger\p_y\psi+V(y)\psi^\dagger\psi\right),}
Clearly, the role of the Planck constant is played by 
$\hbar\equiv 1/(\beta N)$.  

The double-scaling limit of the system corresponds to taking the 
$N\rightarrow\infty$ limit with $\varepsilon_F\rightarrow 0$ while keeping 
$\mu\equiv-N\varepsilon_F$ fixed.  It is convenient to introduce the rescaled 
spatial dimension $\lambda$, 
\eqn\eedoubledim{\lambda=\sqrt{\beta N}\,y.}
After the double-scaling limit, the single-particle equation becomes 
\eqn\eesingleeq{\left(-\frac{1}{2}\frac{\p^2}{\p\lambda^2}+V(\lambda)
\right)\psi(\lambda)=\nu\psi(\lambda),}
where $\nu$ is the double-scaled energy eigenvalue, and 
\eqn\eetwopots{V(\lambda)=\left\{\eqalign{-\frac{1}{2}\omega_0^2\lambda^2\qquad
\quad&\qquad{\rm for\ Type\ 0B},\cr
-\frac{1}{2}\omega_0^2\lambda^2+\frac{(q^2-\frac{1}{4})}{\lambda^2}
&\qquad{\rm for\ Type\ 0A}.\cr}\right.}
The careful definition of the double-scaling limit involves introducing 
a nonuniversal stabilizing regulator $\Lambda$, which we will represent by 
cutting off the potential by an infinite wall at $y\sim 1$.  In the 
double-scaled variable $\lambda$, this amounts to placing an infinite wall at 
$\lambda=\sqrt{2\Lambda}\sim\sqrt{N}$.%
\foot{For a clear discussion of the technical details of the double-scaling 
limit, see, \eg , \kazakov .}

Since $\hbar$ is proportional to $1/N$, the large $N$ limit that we are 
interested in corresponds to the semiclassical limit of the system.  
In the WKB approximation, the semiclassical fermions occupy a certain area in 
phase space, and we have
\eqn\eefirst{N=\int\frac{dp\,d\lambda}{2\pi\hbar}\theta\left(\varepsilon_F
-\frac{p^2}{2}-V(\lambda)\right).}

In a given static vacuum state, one of the main quantities of interest to 
calculate is the vacuum energy
\eqn\eegammadef{F=\lim_{\CT\rightarrow\infty}\left(-\frac{1}{\CT}\log\CZ
\right),}
with $\CT$ is the total length of the Euclidean time dimension. 
In the limit of $\CT\rightarrow\infty$, this is reduced to the evaluation of 
the energy of the ground state, 
\eqn\eeenergy{F=\frac{E_0}{\hbar}=\frac{1}{\hbar}\sum_{k=1}^N\nu_k,}
the sum being performed up to the Fermi energy $N\varepsilon_F\equiv\nu_N$.  
In the double-scaling limit, $F$ represents (a nonperturbative completion of) 
the string partition function, and can be expanded to match the perturbative 
sum over all worldsheet topologies, \ie , over all genera of connected Riemann 
surfaces.  It can be exactly evaluated by first defining the density of states 
$\rho(\mu)$, 
\eqn\eedensstrdef{\rho(\mu)=\hbar\sum\delta(-\mu-\nu_n),}
and observing that in terms of $\rho(\mu)$, we have 
\eqn\eedeltaen{\frac{\p F}{\p\Delta}=\frac{1}{\pi}\mu,\qquad\quad
\frac{\p\Delta}{\p\mu}=\pi\rho(\mu).}
Here $\Delta$ is another scaling variable, usually referred to in the matrix 
models of noncritical strings as the ``worldsheet cosmological constant.'' 
The logarithmic scaling $\rho(\mu)\sim\log\mu$ is a signature behavior of 
two-dimensional string theory \kazmig .  With the use of \eedeltaen , this 
behavior implies for the expansion of $F$ in the powers of the string coupling 
$g_s\sim1/\mu$
\eqn\eestreapprox{F(\mu)\sim\mu^2\ln\mu+\ln\mu+\CO(1/\mu^2).}
The log terms come from the leading $\log\mu$ behavior of the density of 
states, and are characteristic of noncritical string theory in two dimensions 
(in the linear dilaton background, screened by the Liouville wall).  The 
string coupling is determined via $\mu\sim g_s^{-1}$, and the two terms have 
a clear interpretation:  While the first one is the tree-level contribution 
from worldsheets of spherical topology, the second term is a one-loop 
contribution from the torus.  The $\log\mu$ term -- or, more exactly, 
$\log(\Lambda/\mu)$ with $\Lambda$ the cutoff -- is properly interpreted as 
the volume of the Liouville dimension.  

The theory is nonperturbatively fully defined via its free-fermion 
formulation.  A nonlocal transform maps the eigenvalue coordinate to the 
physical spacetime, in which the systems can be understood in terms of 
a spacetime effective theory of strings.  However, this transformation only 
exists under special circumstances, when the $N$ fermions are distributed 
such that the quantum state of the Fermi system can be bosonized in terms of 
hydrodynamic degrees of freedom, such as the fluctuations of the Fermi 
surface.  These fluctuations then correspond in the physical spacetime picture 
to the massless tachyon (and, in Type 0B, the RR scalar) of noncritical 
string theory.  

\subsec{Introducing Noncritical M-Theory}

The spectrum of noncritical Type 0A string theory contains stable D0-branes, 
which couple to a RR one-form gauge field.  It admits vacua with a nonzero 
value of the RR flux $q$.  This flux can also be interpreted as the net number 
of D0-branes sustaining the background.  In the matrix model, $q$ is 
represented as the difference between the number of rows and columns of $M$. 

In the critical Type IIA superstring, stable D0-branes are interpreted as 
KK momentum modes along a hidden, eleventh dimension of M-theory.  It is 
natural to ask whether a similar interpretation can be found for the stable 
D0-branes of the noncritical Type 0A theory, perhaps leading to a noncritical 
version of M-theory in $2+1$ dimensions.  This question can be addressed 
from several points of view.  For example, one can try to identify the lift 
of the effective spacetime action of Type 0A theory to an effective theory in 
$2+1$ dimensions.  Alternatively, one can search for an implementation of the 
lift to M-theory directly in the matrix model.  In this paper, we will 
circumvent some apparent difficulties with these two approaches, by addressing 
the question directly in the language of the second-quantized double-scaled 
fermions.  

It has been observed in \newhat\ that the eigenvalue coordinate $\lambda$ of 
the Type 0A matrix model can be thought of as the radial coordinate on a 
two-dimensional plane (which we will refer to as the ``eigenvalue plane'' from 
now on).  From this viewpoint, the Type 0A vacuum at fixed RR flux $q$ can be 
interpreted as the sector with fixed angular momentum $J=q$ in a $2+1$ 
dimensional theory of fermions on the eigenvalue plane.  Since one unit of the 
D0-brane charge corresponds to one unit of the angular momentum, this leads us 
to a natural lift of the Type 0A vacua to M-theory: 

\medskip
{\it We propose to identify the extra dimension of noncritical M-theory 
with the angular variable on the eigenvalue plane of the double-scaled 
nonrelativistic Fermi system in the upside-down harmonic potential}. 
\medskip

The remainder of this paper can be viewed as a series of tests justifying 
this definition of noncritical M-theory and its proposed relation to 
the dynamics of noncritical strings.  

\subsubsec{A parable on the relation between the radius and the string 
coupling}

At first, the proposed identification of the third dimension of M-theory with 
the angular dimension on the eigenvalue plane may seem somewhat 
counterintuitive.  It suggests that the weakly coupled region in Type 0A 
string theory is associated with the region where the radius of the 
angular $S^1$ dimension of noncritical M-theory is large; similarly, 
the strongly coupled regime of string theory corresponds to the region near 
the origin on the eigenvalue plane where the radius of the angular $S^1$ is 
small.  In contrast, critical M-theory in eleven dimensions relates the strong 
string coupling regime to the large extra dimension of M-theory.  

In order to illustrate that the intuition based on eleven-dimensional M-theory 
may be incorrect in low enough dimensions, consider the following parable, 
which begins with the Einstein-Hilbert action in $D$ spacetime mensions 
$X^\mu$, 
\eqn\eeone{S=\frac{1}{\CG_D}\int d^DX\sqrt{G}\,\CR(G),}
with $G_{\mu\nu}$ the spacetime metric and $\CG_D$ the Newton constant.  
We compactify to $D-1$ dimensions on $S^1$, parametrized by coordinates 
$(X^\mu)=(x^i,Y)$, $i=1,\ldots D-1$, with $Y=Y+2\pi$.  The metric can be 
decomposed as
\eqn\eered{G_{\mu\nu}dX^\mu dX^\nu=e^{2a\Phi}g_{ij}dx^i dx^j +
e^{2b\Phi}dY^2,}
where $\Phi$ is a scalar field (to be identified with the string theory 
dilaton), $g_{ij}$ is the (string frame) metric in $D-1$ spacetime 
dimensions, and $a$ and $b$ are constants to be determined below.  We shall 
only keep the zero modes of all fields on $S^1$, and for simplicity also 
drop the off-diagonal, Abelian gauge field part of $G_{\mu\nu}$.  
Using this decomposition \eered , the Einstein-Hilbert action \eeone\ becomes 
\eqn\eetwo{S=\frac{2\pi}{\CG_D}\int d^{D-1}x\sqrt{g}\left(e^{[(D-3)a+b]\Phi}
R(g)+\ldots\right),}
where ``$\ldots$'' refer to terms that depend on the derivatives of $\Phi$, 
and $R(g)$ is the scalar curvature of the lower-dimensional metric $g_{ij}$.  

If \eetwo\ is to be the leading term of the effective string-theory action in 
the string frame, with $\Phi$ the conventionally normalized dilaton 
(\ie , $e^\Phi=g_s$), the power of $e^\Phi$ in \eetwo\ must equal 
$-2$, implying
\eqn\eethree{(D-3)a+b=-2,\qquad g_s=e^\Phi,\qquad R_D=e^{b\Phi},}
where the third relation -- between the radius $R_D$ of the extra dimension 
measured in the $D$-dimensional Planck units and the dilaton -- follows 
from \eered .  When $D=3$, the first equation in \eethree\ implies that 
$b=-2$, independently of the value of $a$.  Generally, one more 
relation is needed to determine the value of $a$; this extra relation could 
for example come from the requirement that the kinetic term of $\Phi$ be 
correctly normalized, or from a different constraint.  In any case, 
in $D=3$ we do not need to know $a$ to make our point:  Since $b=-2$ 
in $D=3$, the second and third relation in \eethree\ imply that 
the size of the third dimension, measured in the three-dimensional Planck 
units, comes out inversely proportional to the square of the string coupling, 
\eqn\eeradcoup{R_3\sim\frac{\CG_3}{g_s^2}.}
Thus, we see that in the reduction of the simple Einstein-Hilbert Lagrangian 
from three to two dimensions, the large radius of the extra dimension of 
M-theory corresponds to the {\it weak\/} string coupling constant, while the 
strong string coupling regime is described by the {\it small\/} radius of 
the M-theory dimension.  This may be counterintuitive from the viewpoint of 
the critical M-theory in eleven dimensions, but seems compatible with the 
possibility of interpreting the third dimension of noncritical M-theory as 
the angular dimension on a plane.  

Of course, our simple parable has at least two caveats: First of all, the 
eigenvalue plane should not be directly identified with the physical 
spacetime.  Instead, they should be related by a nonlocal transform analogous 
to the tranform between the eigenvalue dimension and the Liouville dimension 
in noncritical string theory. Secondly, the full effective action of 
noncritical M-theory in the physical three-dimensional spacetime is likely to 
be much more complicated than the simple Einstein-Hilbert Lagrangian 
considered in the parable.  

\newsec{Nonperturbative M-Theory as a Double-Scaled Fermi Liquid}

Now we can systematically develop the theory from first principles, and 
check that it leads to sensible results.  

We start with a nonrelativistic spinless Fermi field 
$\hat\Psi(t,y_1,y_2)$ in $2+1$ dimensions, before double scaling.  In the 
double scaling limit, $\hat\Psi$ turns into a double-scaled Fermi field 
$\Psi(t,\lambda_1,\lambda_2)$, described by the action
\eqn\eenonrelbef{S_M=\int dt\,d^2\lambda\,\left(i\Psi^\dagger
\frac{\p\Psi}{\p t}-\frac{1}{2}\sum_{i=1,2}\frac{\p\Psi^\dagger}{\p\lambda_i}
\frac{\p\Psi}{\p\lambda_i}+\frac{1}{2}\omega_0^2\sum_{i=1,2}\lambda_i^2
\Psi^\dagger\Psi+\ldots\right).}
Here the ``$\ldots$'' stand for nonuniversal regulating and stabilizing terms 
in the potential.  We will represent them by an infinite wall placed at 
$\lambda=\sqrt{2\Lambda}/\omega_0$.  In the units where $\hbar$ is 
dimensionless, the basic variables $\omega_0$, $t$, $\lambda_i$, and the 
momentum $p_i$ conjugate to $\lambda_i$ have dimensions 1, $-1$, $-1/2$ and 
1/2, respectively.  Until further notice, we will Wick rotate $t$ and 
interpret it as the Euclidean time coordinate.  

\subsec{First Thoughts on the Double-Scaling Limit}

The double-scaling limit has two ingredients, which are not always clearly 
separated in the studies of two-dimensional string theory.  Both steps are 
performed simultaneously, but the first step is more universal while the 
second one is specific to a given solution.  

\item{(1)} Eliminate the nonuniversal features of the potential,  
represented by the cutoff dependence, and take the large-$N$ limit;

\item{(2)} Choose a state, \ie , a distribution of $N$ fermions among 
the available states, whose double-scaling limit is taken.  Identify the 
scaling variable to be held fixed as $N\rightarrow\infty$.  Typically, the 
scaling variable is a combination of $N$ and a conserved quantity such as 
the energy of the Fermi surface or its angular momentum.%
\foot{We define the Fermi surface more generally as the boundary between the 
filled and empty regions in phase space.}

Some simple modifications of this process can be easily implemented, one 
example being the situation when we do not hold the number of fermions $N$ 
fixed, but instead fix a chemical potential.  We will not distinguish such 
modifications from our prescription.

\subsec{Quantum Mechanics of the Double-Scaled Fermi Liquid}

The theory can be easily quantized.  There are two useful representations.  
In the first one, we use the Cartesian coordinates $\lambda_i$, and view 
the system as two decoupled upside-down harmonic oscillators.  In this 
representation, the second-quantized Fermi field $\Psi$ can be expanded 
in terms of products of Type 0B wavefunctions as follows,
\eqn\eefirstexposc{\Psi(t,\lambda_i)=\int d^2\nu\sum_{s_1,s_2=\pm}a_{s_1s_2}
(\nu_1,\nu_2)\,
\psi_{s_1}(\nu_1,\lambda_1)\psi_{s_2}(\nu_2,\lambda_2)\,e^{-i(\nu_1+\nu_2)t},}
where $\nu_i$ are the energy levels of the two one-dimensional upside-down 
oscillators, and $s_i=\pm$ are the parity quantum numbers of the Type 0B 
wavefunctions.  The annihilation operators $a_{s_1s_2}(\nu_1,\nu_2)$ 
and their conjugates satisfy the canonical anticommutation relations,
\eqn\eefirstantiosc{\{a_{s_1s_2}(\nu_1,\nu_2),a^\dagger_{s_1's_2'}(\nu_1',
\nu_2')\}=\delta_{s_1s_1'}\delta_{s_2s_2'}\delta^2(\nu_i-\nu_i').}

Alternatively, we can use a representation in terms of polar coordinates 
$\lambda,\theta$ on the eigenvalue plane, expanding $\Psi$ in a complete 
basis of Type 0A wavefunctions 
\eqn\eesecondexposc{\Psi(t,\lambda_i)=\sum_{q\in\Z}e^{iq\theta}\int d\nu\,
a_q(\nu)\,\psi_q(\nu,\lambda)\,e^{-i\nu t},}  
supplemented with the canonical commutation relations
\eqn\eesecondantiosc{\{a_q(\nu),a^\dagger_{q'}(\nu')\}=\delta_{qq'}\delta(
\nu-\nu').}
In these formulas, $q$ is the value of the Type 0A RR flux, interpreted in the 
M-theory context as the angular momentum 
on the eigenvalue plane.  The Type 0A and 0B wavefunctions $\psi_{s_1s_2}
(\nu_1,\nu_2)$ and $\psi_q(\nu)$ are given explicitly in terms of cylindric 
Whittaker functions \refs{\gmrev,\moore,\demetrod}.  

\subsec{The Moduli Space of Solutions}

The simplicity of the quantum mechanics of the double-scaled Fermi system 
allows us to make some general remarks about the space of all solutions of 
noncritical M-theory.  These observations will be illustrated in specific 
examples in the rest of the paper.  

In the double scaling limit, the nonuniversal anharmonic pieces in the 
potential are scaled away, and the double-scaled Fermi theory becomes free. 
This leads to a particularly simple description of all possible quantum states 
in this theory.  In order to specify a quantum state $|{\rm phys}\rangle$, we 
simply need to decide how each canonical pair of oscillators $a,a^\dagger$ 
acts on $|{\rm phys}\rangle$.  Any quantum state that can be prepared 
by the infinite collection of fermionic oscillators is a solution of 
noncritical M-theory.  Most such states will not have a clear semiclassical 
description in terms of collective bosonic degrees of freedom, since generally 
each canonical pair can act on $|{\rm phys}\rangle$ in a way uncorrelated with 
the action of the other pairs.  Only those states for which the fermionic 
oscillators act in a highly correlated way will exhibit semiclassical 
hydrodynamic bosonic excitations. We will refer to such states generically 
as ``hydrodynamic states.'' We expect that the excitations of such a 
hydrodynamic state can be described in terms of an effective action 
for the fluctuations of the hydrodynamic bosonic variables (such as the 
bosonic fluctuations of the Fermi surface, whenever the latter can be 
defined).  Only states that can be so bosonized can be described in terms of 
low-energy quantum gravity in a semiclassical spacetime.  The physical 
spacetime itself is an emergent property of the hydrodynamic states, and is 
related in a complicated nonlocal way to the eigenvalue plane on which the 
fermions reside.  

It is worth noting that our definition of noncritical M-theory in terms 
of free fermions leads to a very precise refinement of the famous ``starfish 
diagram,'' traditionally drawn to illustrate the space of all vacua in  
critical string/M theory.  This starfish diagram usually depicts several 
asymptotic corners, in which perturbative string/M-theory descriptions are 
available, connected into a single moduli space whose middle portion remains 
rather mysterious. In contrast, as we have just argued, the problem of 
identifying the space of all solutions in our noncritical M-theory is 
effectively reduced to a simple, mathematically well-posed problem, 
essentially equivalent to the representation theory of the algebra 
of the inifinite set of decoupled canonical fermionic oscillators 
$a_q^\dagger(\nu)$ and $a_q(\nu)$.  

In this picture, the spacetime effective field theory description is 
effectively equivalent to the hydrodynamics of the Fermi liquid.  Whether or 
not an effective bosonization to a spacetime description exists, however, 
the physics of any given solution is always nonperturbatively fully defined 
by the underlying fundamental degrees of freedom of noncritical M-theory, the 
double-scaled nonrelativistic free fermions on the eigenvalue plane.  
Different quantum vacua of the system correspond to different separable 
Hilbert spaces that can be built as fermionic Fock spaces from a given ground 
state.  This leads to an intricate picture of a web of Hilbert spaces, 
representing all possible ways in which the $N$ fermions can occupy the 
available single-particle states while the double-scaling limit is taken.  
Some such states represent static vacuum solutions, others will describe 
excited states in such vacua (\ie , they belong to the Hilbert space for 
which the corresponding vacuum state is the ground state).   Some solutions 
will be time dependent, interpolating between different static vacua at early 
and late times.  Yet others may represent big-bang/big-crunch cosmologies, 
evolving from/to M-theory states with no conventional semiclassical spacetime 
interpretation.  Some may have a $2+1$-dimensional spacetime, some reduce to 
string vacua in a $1+1$-dimensional spacetime.  Some solutions will have a 
dynamically changing spacetime dimension, evolving for example from $1+1$ at 
early times via $2+1$ at intermediate times to $1+1$ at late times, etc.  
Some simple examples of such classes of solutions will be discussed below, 
but many more can be identified and studied within this rich and 
mathematically well-defined ``landscape of all vacua'' of noncritical 
M-theory.  

\newsec{Examples of Solutions I: Type 0A and 0B Strings from M-Theory}

As a first check that our definition of noncritical M-theory is acceptable, 
we shall reproduce known Type 0A and 0B vacua as its solutions. 

\subsec{Type 0A}

Using the polar-coordinate representation of the theory, we first choose a 
value of the RR flux $q$, and define the Type 0A state $|\,{\rm 0A},q,\mu
\rangle$ 
as a solution of noncritical M-theory, as follows.  The $N$ fermions are 
distributed such that the Fermi sea is filled up to some (negative) Fermi 
energy $-\mu$ in the sector with angular momentum $q$ while keeping the Fermi 
sea empty in all the sectors with angular momenta $q'\neq q$: 
\eqn\eedeftypeoa{\eqalign{a_q(\nu)\,|\,{\rm 0A},q,\mu\rangle&=0\qquad{\rm 
for}\ \nu>-\mu,\cr
a_q^\dagger(\nu)\,|\,{\rm 0A},q,\mu\rangle&=0\qquad{\rm for}\ \nu<-\mu,\cr
a_{q'}(\nu)\,|\,{\rm 0A},q,\mu\rangle&=0\qquad{\rm for\ all}\ \nu\ {\rm with}\ 
q'\neq q.\cr}}
Notice that it is important to use this definition while taking the double 
scaling limit.  In particular, this state is not equivalent to sending 
$\mu\rightarrow\infty$ in all sectors with $q'\neq q$ after the double scaling 
limit has been performed.  To see this, we shall now reproduce the known 
result for the exact vacuum energy of the Type 0A solution, from a direct 
M-theory calculation.    

The total vacuum energy of the 0A state $|\,{\rm 0A},q,\mu\rangle$ will be 
equal to the sum of vacuum energies over all M-theory sectors of fixed angular 
momentum $q$, filled up to a $q$-dependent Fermi level as indicated in 
\eedeftypeoa .  The naive limit $\mu\rightarrow\infty$ in sectors of 
$q'\neq q$ is properly interpreted as a prescription to keep the Fermi level 
at the cutoff $\Lambda$ during the double scaling limit.  Recall that at fixed 
$q$, the density of states in Type 0A theory has an asymptotic string-coupling 
expansion \refs{\newhat,\deformed-\demklerod}%
\foot{Throughout the paper, we use the ``$\,\approx\,$'' symbol to denote 
exact asymptotic expansions, reserving ``$\,\sim\,$'' to represent a more 
loosely defined proportionality or scaling relation.}
\eqn\eereproenoa{\rho_{0A}(\mu,q)\approx-\frac{1}{4\pi}\log(\mu^2+q^2)
+\CO(1/\mu^2).}
Sending formally $\mu\rightarrow\infty$ would kill all the terms 
$\CO(1/\mu)$ and higher, but we would still be left with the leading log.  
Keeping track of the cutoff dependence in $\rho_{0A}$ during the double 
scaling, we find an extra correction $\sim\frac{1}{2\pi}\log\Lambda$ in the 
density of states (see, \eg , \kazakov).  Thus, setting $\mu=\Lambda$ in all 
sectors $q'\neq q$ and taking the double scaling limit $\Lambda\rightarrow
\infty$ will eliminate also the log contribution from all sectors of 
$q'\neq q$, and the resulting density of states of this M-theory solution is 
manifestly equal to the density of states in Type 0A theory at RR flux $q$.  
The integration of $\rho$ to obtain the vacuum energy is then 
straightforward.  

\subsec{Type 0B}

The Type 0B linear dilaton vacuum $|\,{\rm 0B},\mu\rangle$ can be defined as 
a solution of noncritical M-theory as follows.  In the Cartesian 
representation of the theory, the energies $\nu_1$ and $\nu_2$ of the two 
one-dimensional oscillators are separately conserved.  Fix all the quantum 
numbers of the second oscillator, \ie\ pick an arbitrary fixed 
value $\bar\nu_2$ of $\nu_2$ and $\bar s_2$ of $s_2$, and fill all states in 
the sector with $\nu_2=\bar\nu_2$ and $s_2=\bar s_2$ up to a negative Fermi 
energy $-\mu$ while keeping the Fermi sea empty in all sectors with 
$\nu_2\neq\bar\nu_2$ or $s_2\neq\bar s_2$:

\eqn\eedeftypeob{\eqalign{a_{s_1s_2}(\nu_1,\nu_2)\,|\,{\rm 0B},\mu\rangle&=0
\qquad{\rm for}\ \nu_1>-\mu\ {\rm with}\ \nu_2=\bar\nu_2\ {\rm and}
\ s_2=\bar s_2,\cr
a_{s_1s_2}^\dagger(\nu_1,\nu_2)\,|\,{\rm 0B},\mu\rangle&=0\qquad{\rm for}
\ \nu_1<-\mu\ {\rm with}\ \nu_2=\bar\nu_2\ {\rm and}\ s_2=\bar s_2,\cr
a_{s_1s_2}(\nu_1,\nu_2)\,|\,{\rm 0B},\mu\rangle&=0\qquad{\rm for\ all}\ 
\nu_1\ {\rm with}\ \nu_2\neq\bar\nu_2\ {\rm or}\ s_2\neq\bar s_2.\cr}}
Several observations:

\item{$\bullet$} Unlike in Type 0A, selecting a fixed value $\nu_2=\bar\nu_2$ 
to fill the Fermi sea does not introduce any new physical free parameter, as 
any change in the value of $\bar\nu_2$ can be absorbed in a shift of $\mu$.  
Without any loss of generality, we can take $\bar\nu_2=0$.  

\item{$\bullet$} The parallels between the Type 0A and Type 0B constructions 
are even stronger before the double-scaling limit.  In that situation, $\nu_2$ 
is also a discrete conserved quantum number. 

\item{$\bullet$} The bosonic $c=1$ string can also be easily found as a 
solution of noncritical M-theory, by repeating the steps of our Type 0B 
construction and filling only one side of the one-dimensional effective 
potential at fixed $\nu_2=\bar\nu_2$ and $s_2=\bar s_2$. 

\item{$\bullet$} Similarly, our construction can be easily extended to simple 
orbifolds of Type 0A and 0B theories, such as the IIA and IIB models 
considered in \refs{\gtato,\takaiia,\seibergiia}.%
\foot{We also mention in passing that a duality diagram has been proposed 
some time ago for critical 0A and 0B in ten dimensions in \oren , 
conjecturally connecting them to nonsupersymmetric compactifications of 
M-theory.  Our results do not have any direct bearing on whether or 
not the proposal of \oren\ is correct.  Unlike Type 0 theories in the critical 
dimension, the two-dimensional models that we consider do not suffer from 
instabilities, and the duality properies are thus under control, and amenable 
to our exact analysis.}
\item{$\bullet$} The quantum states defining the Type 0A and 0B theories 
exhibit a semiclassical Fermi surface which is effectively of higher 
codimension in phase space, compared to the naive ground state of the system 
(to which we return in Section~5).  This is somewhat reminiscent of 
the higher-codimension Fermi surfaces classified and related to K-theory in 
\kfermi , the main difference being that the system of spinless fermions is 
not in the stable regime of K-theory.  Defining the proper semiclassical limit 
of such states at large $N$ might require the more systematic approach to 
large $N$ developed in \yaffe .  

\newsec{Examples of Solutions II: The M-Theory Vacuum in 2\/+\/1 Dimensions}

In critical M-theory, perhaps the most interesting vacua are those that 
exhibit the largest spacetime symmetry in uncompactified eleven dimensions: 
The flat Minkowski space, described at low energies by eleven-dimensional 
supergravity, and the heterotic M-theory solution \hw, with 
the additional $E_8$ super Yang-Mills at the boundary of spacetime.  It is in 
those solutions where the non-stringy character of M-theory is most prominent, 
since neither of these two vacua admits string-like excitations.  
Having reproduced the two-dimensional string theory vacua from our noncritical 
M-theory in the previous subsection, we can now analyze its ``non-stringy 
phase,'' and in particular, its $2+1$ dimensional vacua. 

The noncritical M-theory has one particularly natural solution, corresponding 
to filling the states up to some common Fermi energy $\varepsilon_F$ in the 
$2+1$ dimensional system of fermions, irrespective of their other quantum 
numbers.  We will refer to this solution as the ``M-theory vacuum.''  By 
construction, it represents the M-theory lift of the linear dilaton vacua 
of Type 0A and 0B theories.  Thus, we define the M-theory vacuum state 
$|\,{\rm M},\mu\rangle$ -- using, for definiteness, the polar-coordinate 
representation of the theory -- as follows:
\eqn\eepoldefm{\eqalign{a_q(\nu)\,|\,{\rm M},\mu\rangle&=0\qquad{\rm for}\ 
\nu>-\mu\ {\rm and\ all}\ q,\cr
a_q^\dagger(\nu)\,|\,{\rm M},\mu\rangle&=0\qquad{\rm for}\ \nu<-\mu\ {\rm and\ 
all}\ q.\cr}}
Strictly speaking, one should distinguish between the definition of the 
M-theory state before and after the double scaling limit. However, we shall 
keep the distinction implicit, in order to keep the notation simple.  
We shall now analyze the scaling properties of this state, in order to 
identify appropriately the double-scaling limit of $|\,{\rm M},\mu\rangle$. 

\subsec{Scaling at Leading Order in $1/N$}

The large $N$ limit corresponds to the WKB approximation of the M-theory 
vacuum defined in \eepoldefm .  In this limit, the semiclassical density of 
states is given by
\eqn\eeleadrho{\rho(\nu)=\hbar\int\frac{d^2p\,d^2 y}{(2\pi\hbar)^2}\,
\delta\left(\nu-h(p_i, y_i)\right),}
where the single-particle Hamiltonian is 
\eqn\eesingham{h(p_i,y_i)=\frac{1}{2}\sum_{i=1,\,2}(p_i^2-\omega_0^2
y_i^2+\ldots),}
where we will only keep track of the universal part in the potential.  
Introducing $y=\sqrt{y_1^2+y_2^2}$ and $p=\sqrt{p_1^2+p_2^2}$, and switching 
to the polar coordinates separately in the coordinate and momentum space, 
gives
\eqn\eeleadrhoa{\rho(\nu)=\int \frac{dp\,dy}{\hbar} py\,
\delta\left(\nu-\frac{1}{2}p^2+\frac{1}{2}\omega_0^2y^2\right).}
We will use a rotationally invariant cutoff $\Lambda$, equivalent to placing 
an infinite wall at $\lambda\leq\sqrt{2\Lambda}/\omega_0$.  The integration 
gives
\eqn\eeleadrhob{\rho(\nu)=\frac{1}{\hbar}\int_{\sqrt{-2\varepsilon_F}/
\omega_0}^{\sqrt{2\Lambda}/\omega_0}y\,dy\sim\frac{1}{\hbar\omega_0^2}
(\varepsilon_F+\Lambda)\sim\frac{\varepsilon_F}{\hbar\omega_0^2},}
where in the final step we have dropped the nonuniversal cutoff-dependent 
part of the density of states, keeping only its dependence on 
$\varepsilon_F$.  

We can use this evaluation of the density of states to obtain an expression 
for the vacuum energy of the system.  In the semiclassical regime, each 
fermion occupies a unit volume $1/(2\pi\hbar)^2$ in phase space, and the total 
number $N$ of fermions measures the semiclassical area of the filled region,
\eqn\eeleadn{N(\varepsilon_F)=\int\frac{d^2p\,d^2y}{(2\pi\hbar)^2}\,
\theta(\varepsilon_F-h(p_i,y_i)).}
Similarly, the semiclassical ground-state energy is given by
\eqn\eeleade{E_0(\varepsilon_F)=\int\frac{d^2p\,d^2y}{(2\pi\hbar)^2}\,
h(p_i,y_i)\,\theta(\varepsilon_F-h(p_i,y_i)).}
Taking the derivative of each of those equations with respect to 
$\varepsilon_F$, we get
\eqn\eeleadnder{\frac{\p N(\nu)}{\p\nu}=\int\frac{d^2p\,d^2y}{(2\pi
\hbar)^2}\,\delta(\nu-h(p_i,y_i))}
and
\eqn\eeleadeder{\frac{\p E_0(\nu)}{\p\nu}=\int\frac{d^2p\,d^2y}{(2\pi
\hbar)^2}\,h(p_i,y_i)\,\delta(\nu-h(p_i,y_i))=\nu\frac{\p N}{\p\nu}.}
Since the density of states is related to $N$ via
\eqn\eedensidef{\rho(\nu)=\hbar\frac{\p N}{\p\nu},}
 we finally obtain
\eqn\eeleadf{\eqalign{F(\varepsilon_F)\equiv\frac{E_0(\varepsilon_F)}{\hbar}
&=\frac{1}{\hbar}\int^{\varepsilon_F}d\nu\,\nu\frac{\p N}{\p\nu}=
\frac{1}{\hbar^2}\int^{\varepsilon_F}d\nu\,\nu\rho(\nu)\cr
&\sim\frac{1}{\hbar^3\omega_0^2}\left(\frac{\varepsilon_F^3}{3}+
\frac{\varepsilon_F^2\Lambda}{2}\right)\sim\frac{\varepsilon_F^3}{3
\hbar^3\omega_0^2}.\cr}}
In the final step, we have again kept only the universal dependence on 
$\varepsilon_F$.%
\foot{From now on, we set $\beta=1$ by rescaling the corresponding variables.  
Hence, $\hbar=1/N$.}

\smallskip
From this central result, we can draw several interesting lessons:  
\smallskip

\item{(1)} The natural scaling variable suggested by this lowest-order 
result is 
\eqn\eenatscaling{\mu=-\varepsilon_F/\hbar\equiv -N\varepsilon_F.}
It is satisfying to find that the scaling variable in M-theory is indeed 
the same as in Type 0A and 0B theory.  
\smallskip

\item{(2)} The vacuum energy $F\sim\int\nu\rho(\nu)\,d\nu$ scales as 
\eqn\eenatescaling{F\sim-\mu^3+\ldots.} 
This behavior seems characteristic of M-theory.  In the physical spacetime 
interpretation of this result, the leading term in $F$ should 
correspond to the tree-level contribution, proportional to $\kappa^{-2}$ where 
by $\kappa$ we denote the spacetime coupling constant, possibly related to the 
Newton constant.  This implies that in terms of $\kappa$, the natural 
expansion parameter $1/\mu$ of noncritical M-theory is 
\eqn\eenatexpansion{\frac{1}{\mu}=\kappa^{2/3}.}
This, of course, is the behavior observed in critical heterotic M-theory \hw.  
It is also suggestive of a possible existence of membranes in the noncritical 
M-theory.  
\smallskip

\item{(3)} Viewed in the context of large $N$ theories, our $2+1$ dimensional 
model exhibits an interesting behavior: At the leading order at large $N$, 
our vacuum energy scales as 
\eqn\eescalefn{F\sim N^3,}
to be contrasted with the more conventional $F\sim N^2$ behavior familiar 
from the traditional ``planar'' large $N$ limit.    
\smallskip

\item{(4)} Unlike in string theory in $1+1$ dimensions, there is no 
logarithmic dependence of the density of states on the Fermi energy.  In 
noncritical strings, such a logarithmic dependence signifies the volume 
dependence of various terms in the sum over surfaces; its absence here 
suggests that the dependence on volume is reduced in M-theory.  We shall 
return to this point in Section~6.3, where the volume/cutoff dependence of 
the M-theory amplitudes will be discussed. 
\smallskip

\item{(5)} As in noncritical string theory in $1+1$ dimensions, the system 
exhibits a particle-hole duality, accompanied by the exchange%
\foot{This is an important symmetry, since it is related in the string theory 
context to the orbifold that produces Type IIA and IIB out of Type 0A and 0B 
vacua \refs{\gtato,\takaiia,\seibergiia}.  It is satisfying to see that a 
similar symmetry, and hence an orbifold procedure, extends to noncritical 
M-theory.}
\eqn\eeexch{\mu\rightarrow -\mu.}
Notice that the Fermi surface undergoes a topology changing transition as 
$\mu$ goes from positive to negative values, but the geometry of the Fermi 
surface afer the transition is the same as its geometry before the transition. 
As a result of this nonperturbative symmetry we expect that $\rho$ (and 
consequently $F$) should be an even function of $\mu$.  Surprisingly, this 
expectation is apparently violated by the leading scaling behavior of $\rho$ 
and $F$ that we just determined in the WKB approximation.  This apparent 
paradox will be resolved when we obtain the exact nonperturbative formula 
for $\rho$, which will indeed be an even function.  The odd piece in the 
perturbative expansion is a consequence of the expansion in powers of 
$1/\mu$, which splits the exact formula into a perturbative and a 
nonperturbative piece, neither of which is even under \eeexch .

\subsec{The Double-Scaling Limit}

Having identified the correct scaling variable, it is now clear how to define 
the double-scaling limit of the M-theory vacuum.  It is given by distributing 
$N$ fermions such that they fill all the lowest energy levels up to a Fermi 
energy $\varepsilon_F$, and then taking the limit 
\eqn\eedoublemlim{N\rightarrow\infty,\qquad\varepsilon_F\rightarrow 0,\qquad 
\mu\equiv -N\varepsilon_F\ \ {\rm fixed}.}
Thus, the rules for taking the double scaling limit of fermions in the 
M-theory vacuum turn out to be exactly the same as in noncritical string 
theory in $1+1$ dimensions.  The double-scaling limits leading to Type 0A, 
0B or M-theory solutions differ only in the selection of how the $N$ fermions 
occupy available energy levels, but not in how the $N\rightarrow \infty$ limit 
is taken.  

\subsec{The Worldsheet Cosmological Constant and the String Susceptibility}

If this were a string theory, we would be interested in expressing the 
amplitudes in terms of the worldsheet cosmological constant $\Delta$.  
In the matrix model, this cosmological constant can be defined via  
\eqn\eedeltad{\frac{\p\Delta}{\p\mu}=\pi\rho(\mu).}
In string theory, a particularly important critical exponent is 
$\gamma_{\rm str}$, known as the ``string susceptibility'' exponent 
\refs{\gmrev-\nakarev}. It is usually defined via the leading scaling 
behavior of the vacuum energy of the matrix model in the double scaling limit, 
\eqn\eesuscdef{F\sim\Delta^{2-\gamma_{\rm str}}+\ldots}
In matrix models of noncritical strings, we are limited to backgrounds with 
spacetime dimension $d\leq 2$.  These backgrounds have 
$\gamma_{\rm str}\leq 0$, with the bound $\gamma_{\rm str}=0$ saturated for 
two-dimensional strings, or central charge $c$ (or $\hat c$) equal to one.  
This is the famous ``$c=1$ barrier''  of the matrix model formulation 
of noncritical string theory.%
\foot{The $c=1$ barrier can perhaps be breached by considering supersymmetric 
noncritical strings with an exotic type of supersymmetry, \highnon .}

In noncritical M-theory, we can define $\Delta$ as in \eedeltad .  In the 
leading WKB approximation, we get
\eqn\eedeltam{\Delta\approx-\frac{\pi}{2\omega_0^2}\mu^2+\ldots}
Furthermore, we can introduce the ``M-theory susceptibility'' exponent 
$\gamma_M$, defined exactly as in string theory via \eesuscdef .  As we have 
seen, in the M-theory vacuum $F\sim\mu^3+\ldots$, implying 
\eqn\eemsusc{\gamma_M=1/2.}
Thus, the noncritical M-theory vacuum is in the regime of values of the 
susceptibility exponent that is unattainable by matrix models of noncritical 
strings.  This is yet another check that our M-theory is naturally interpreted 
as living beyond the $c=1$ barrier, at the cost of not being a string theory 
anymore.  

\newsec{Exact Vacuum Energy in Noncritical M-Theory}

Having defined our M-theory vacuum solution $|\,{\rm M},\mu\rangle$, we can 
now use the exact free-fermion description of the system to extract a wealth 
of physical information about $|\,{\rm M},\mu\rangle$ and its excitations.  
As an example, we will evaluate the exact vacuum energy of this solution.  

As in the matrix models of noncritical string theory, the vacuum energy 
$F(\mu)$ is determined in terms of the exact density of states $\rho_M(\mu)$ 
via
\eqn\eevacendens{\frac{\p F}{\p\mu}=\mu\rho_M(\mu).}
The $\mu$ derivative of $\rho_M$ can be expressed in the following, 
cutoff-independent integral representation, 
\eqn\eederden{\eqalign{\frac{\p\rho_M}{\p\mu}=\sum_{q\in\Z}\frac{\p
\rho_{0A}}{\p\mu}&=\frac{1}{2\pi\omega_0\mu}\Im\int_0^\infty d\sigma\,
e^{-i\sigma}\frac{\omega_0\sigma/\mu}{\sinh\{\omega_0\sigma/\mu\}}
\sum_{q\in\Z}e^{-|q|\omega_0\sigma/\mu}\cr
&=\frac{1}{2\pi\omega_0\mu}\Im\int_0^\infty d\sigma\,e^{-i\sigma}
\frac{\omega_0\sigma/(2\mu)}{\sinh^2\{\omega_0\sigma/(2\mu)
\}}.\cr}}
Recall that the scale $\omega_0$ is related in Type 0A and 0B string theory 
to the string scale via $\omega_0=1/\sqrt{2\alpha'}$.  We obtained the 
integral representation \eederden\ by summing the Type 0A contributions 
\refs{\newhat,\demetrod} from the  sectors of all integer values of RR flux 
$q$.  

Alternatively, this integral representation could be obtained in a manner 
closer to Type 0B, using the Cartesian coordinate representation of M-theory 
and the definition of the density of states via the resolvent of the 
one-particle Hamiltonian $h(p_i,\lambda_i)$,
\eqn\eerhoresolv{\rho_M(\mu)=\lim_{\epsilon\rightarrow 0^+}\frac{1}{\pi}\Im
\Tr\left(\frac{1}{h(p_i,\lambda_i)+\mu-i\epsilon}\right).}
The resolvent can be easily evaluated, leading to
\eqn\eeresolex{\eqalign{\langle\widetilde\lambda_i|&\frac{1}{h-\mu-i\epsilon}
|\lambda_j\rangle={}\cr
{}&=i\int_0^\infty d\tau e^{-i\mu\tau}\left(\frac{i\omega_0}{2\pi
\sinh(\omega_0\tau)}\right)\exp\left\{\frac{i\omega_0[(\lambda^2+
\widetilde\lambda^2)\cosh(\omega_0\tau)-2\lambda\widetilde\lambda]}{2\sinh
(\omega_0\tau)}\right\}.\cr}}
Upon evaluating the Gaussian integrals over $\lambda_i$ and using 
\eerhoresolv , we obtain 
\eqn\eedensnice{\rho_M(\mu)=\frac{1}{4\pi}\Re\int_0^\infty d\tau\,
e^{-i\mu\tau}\frac{1}{\sinh^2(\omega_0\tau/2)}.}
Strictly speaking, this formula depends on a cutoff (\ie , is formally 
divergent near $\tau=0$ and needs to be regulated), but in a way which is 
$\mu$ independent.  Taking the derivative of \eedensnice\ and rescaling the 
integration variable to $\sigma=\mu\tau$, we reproduce \eederden .  

\subsec{The Weak Coupling Expansion}

We see that the M-theory vacuum has a dimensionless parameter, 
$\mu/\omega_0$.  In string theory, this parameter would play the role of the 
inverse string coupling constant.  Thus, in analogy with string theory, we 
first study the perturbation expansion in the powers of $1/\mu$.  

\subsubsec{Leading order}

The leading term in the expansion of the density of states is 
$\mu$-independent, and equal to
\eqn\eeleadingind{\eqalign{\frac{\p\rho_M}{\p\mu}&\approx -\frac{1}{\pi
\omega_0^2}\int_0^\infty\frac{d\sigma}{\sigma}\sin \sigma\,e^{-\epsilon\sigma}
+\CO(1/\mu^2)\cr
&=-\frac{1}{2\omega_0^2}+\CO(1/\mu^2).\cr}}
This is to be contrasted with the $1/\mu$ expansion in two-dimensional string 
theory, where the leading term in $\p\rho/\p\mu$ goes as $1/\mu$, leading to 
the characteristic logarithmic behavior of $\rho(\mu)$.  

\vfill\break
\subsubsec{Higher loops}

In the $1/\mu$ expansion of the derivative of the density of states 
\eederden , only even powers of $1/\mu$ appear, and the term of order $2m$ 
(with $m=1,2,\ldots$) is proportional to 
\eqn\eepropzero{\int_0^\infty d\sigma\,\sigma^{2m-1}\sin \sigma\,e^{-\epsilon
\sigma}.}
All such integrals vanish identically, implying that our perturbation series 
for $\p\rho_M/\p\mu$ terminates after the lowest, constant term, and we 
obtain   
\eqn\eehurv{\frac{\p\rho_M}{\p\mu}\approx -\frac{1}{2\omega_0^2}+{\rm 
possible\ nonperturbative\ terms}.}
This in turn leads to the asymptotic expansion of the density of states
\eqn\eedenshurv{\rho_M(\mu)\approx -\frac{1}{2\omega_0^2}\,\mu+\frac{C}{
\omega_0},}
where $C$ is a nonuniversal dimensionless integration constant, to be 
discussed in Section~6.3. Finally, this yields the perturbative formula for 
the vacuum energy, exact to all orders in powers of $1/\mu$, 
\eqn\eevaehurv{F=\int^{-\mu}\nu\rho_M(\nu)d\nu\approx\frac{1}{2\omega_0^2}
\int^{-\mu}\nu^2d\nu=-\frac{1}{6\omega_0^2}\,\mu^3+\frac{C}{2\omega_0}\,\mu^2
+\omega_0C_0.}
The integration constant $C_0$ represents a one-loop term.  Unlike in 
noncritical string theory, where the one-loop term is proportional to 
$\log\mu$, $C_0$ in M-theory is $\mu$ independent, and can therefore be 
eliminated by a shift in the overall zero of energy the vacuum energy $F$.  
We will set $C_0=0$ from now on.  

We found a dramatic simplification in the $1/\mu$ expansion of the 
vacuum energy of the M-theory vacuum, compared to its string theory 
counterparts (where all orders in $1/\mu$ are generically nonzero).  
The fact that the perturbative expansion terminates at one loop is a first 
hint that the theory may be topological, or at least exhibit a localization 
of the path integral similar to that of a topological theory.  This will be 
further confirmed when we study the structure of nonperturbative corrections 
to the vacuum energy below.  

\subsubsec{Summation of higher-genus Type 0A contributions}

From the point of view of string theory, this result can be reproduced by 
summing the asymptotic expansions of the Type 0A amplitudes order by order 
in $1/\mu$.  The vacuum energy in Type 0A theory with RR flux $q$ can also 
be expanded in the powers of the string coupling $1/\mu$, and the coefficients 
of this series are generically nonzero to all orders.  It is instructive to 
see how they sum up to zero, order by order in $1/\mu$, when the summation 
over $q$ is performed.  In the derivative of the density of states, these 
terms of order $1/\mu^3$ and higher are (with the implicit $\zeta$-function 
regularization) 
\eqn\eehigherorder{\eqalign{\frac{\p\rho}{\p\nu}
&=\frac{2}{\pi\omega_0^2}\sum_{k=1}^{\infty}\frac{k^2\nu/\omega_0}{(k^2+\nu^2/
\omega_0^2)^2}\cr
&\approx\frac{2\omega_0}{\pi\nu^3}\sum_{k=1}^\infty\sum_{m=0}^\infty 
(m+1)(-1)^m k^{2m}\left(\frac{\omega_0}{\nu}\right)^{2m}\cr
&\approx-\frac{2}{\pi\omega_0^2}\sum_{m=1}^\infty(-1)^m m\zeta(-2m)
\left(\frac{\omega_0}{\nu}\right)^{2m+1}.\cr}}
Hence, the nonzero contributions from sectors of fixed $q$ sum up, at order 
$(1/\mu)^{2m+1}$, to give $\zeta(-2m)$.  Since $\zeta(-2m)=0$ for $m=1,
\ldots$, all terms $m=1,\ldots$ in this asymptotic expansion are identically 
zero. 

\subsubsec{Summation of the leading logs}

It is similarly instructive to see how the leading $\mu^3$ behavior of the 
vacuum energy in M-theory comes about from the summation of the leading 
$\mu^2\log\mu$ terms in Type 0A at fixed $q$. (For simplicity, we set 
$\omega_0=1$ in this paragraph.)

Recall that at fixed $q\in\Z$, the density of states in Type 0A theory has an 
asymptotic expansion \refs{\newhat,\demetrod,\demklerod}%
\eqn\eedenszeroa{\rho_{0A}(\nu,q)\approx-\frac{1}{2\pi}\Re\log(|q|-
i\nu)+\CO(1/(|q|-i\nu)).}
In M-theory, we fill all sectors with different values of $q$ up to 
the common Fermi level.  The leading term in the expansion of the density of 
states is then
\eqn\eedensmdisc{\rho_M(\mu)=\sum_{q\in\Z}\rho_{0A}
(\mu,q)\approx-\frac{1}{4\pi}\sum_{q\in\Z}\log(\mu^2+q^2)+\ldots}
We are interested in summing these leading logs.  We have
\eqn\eedensmdisclogs{\eqalign{-\frac{1}{4\pi}\sum_{q\in\Z}&\log(\mu^2+q^2)
=-\frac{1}{4\pi}\log(\mu^2)-\frac{1}{2\pi}\sum_{q=1}^\infty
\log\left(q^2(1+\mu^2/q^2)\right)\cr
&=-\frac{1}{4\pi}\log(\mu^2)-\frac{1}{2\pi}\sum_{q=1}^\infty
\log(q^2)-\frac{1}{2\pi}\log\prod_{q=1}^\infty(1+\mu^2/q^2)\cr
&=-\frac{1}{4\pi}\log(\mu^2)-\frac{1}{2\pi}\log\left[\frac{\sinh(\pi\mu)}{\pi
\mu}\right]+\ldots\cr
&=-\frac{1}{2\pi}\log\left(\sinh(\pi\mu)\right)+\ldots
=-\frac{\mu}{2}-\frac{1}{2\pi}\log\left(1-e^{-2\pi\mu}\right)
+\ldots\cr
&=-\frac{\mu}{2}+\ldots\cr}}
where the ``$\ldots$'' in \eedensmdisclogs\ refer to divergent but 
$\mu$-independent terms, and where in the final formula we also dropped all 
the terms nonperturbative in $1/\mu$.   

Thus we see that the leading $\log\mu$ piece from the $q=0$ sector is exactly 
offset by a contribution from $\log\sinh(\pi\mu)/\mu$ which originates in the 
sum over sectors with $q\neq 0$.  Instead, the leading log is replaced by a 
term linear in $\mu$, which also emerges from the sum over all $q$.  
Consequently, we end up with the M-theory scaling, 
\eqn\eemscaling{\rho_M(\mu)\sim\mu\equiv\kappa^{-2/3},}
predicted by the WKB argument of the previous subsection.  

\subsec{The Strong Coupling Expansion}

We now turn to the analysis of the nonperturbative corrections.  

The integral representation for the derivative of the density of states can be 
expanded in the powers of $\mu$:

\eqn\eedermuexp{\eqalign{\frac{\p\rho_M}{\p\mu}&\approx -\frac{1}{\pi
\omega_0^2}\sum_{n=0}^\infty(-1)^n\frac{(2\mu/\omega_0)^{2n+1}}{(2n+1)!}
\int_0^\infty d\tau\frac{\tau^{2n+2}}{\sinh\tau}\cr
&\approx-\frac{1}{\omega_0^2}\sum_{n=0}^\infty\frac{(2\pi\mu/
\omega_0)^{2n+1}B_{2n+2}}{(2n+1)!}.\cr}}
Alternatively, this same result can be obtained by summing Type 0A 
contributions over all values of $q$. Indeed, 
\eqn\eehighoinv{\eqalign{\frac{\p\rho_M}{\p\nu}&=
\frac{2}{\pi\omega_0^2}\sum_{k=1}^\infty\frac{\nu/\omega_0^2}{k^2(1+
\nu^2/(k^2\omega_0^2))^2}\cr
&\approx\frac{2}{\pi\omega_0^2}\sum_{k=1}^\infty\sum_{m=0}^\infty (m+1)(-1)^m
\frac{(\nu/\omega_0)^{2m+1}}{k^{2m+2}}\cr
&\approx-\frac{2}{\pi\omega_0^2}\sum_{m=1}^\infty m(-1)^m\zeta(2m)\left(
\frac{\nu}{\omega_0}\right)^{2m-1}.\cr}}
The zeta function at positive even integers can be expressed in terms of the 
Bernoulli numbers,
\eqn\eebern{\zeta(2n)=\frac{2^{2n-1}\pi^{2n} |B_{2n}|}{(2n)!}.}
Using the fact that $B_{2n}=(-1)^{n+1}|B_{2n}|$ for $n=1,\ldots$, we get
\eqn\eefirstsum{\frac{\p\rho_M}{\p\nu}\approx-\frac{1}{\omega_0^2}
\sum_{m=1}^\infty 2m\,\left(\frac{2\pi\nu}{\omega_0}\right)^{2m-1}
\frac{B_{2m}}{(2m)!},}
reproducing \eedermuexp .

\subsec{Dependence on the Cutoff and Volume}

So far, we have only considered the universal part of the density of states 
and the vacuum energy.  Now we take a closer look at the possible cutoff 
dependence.  Recall that in noncritical string theory, the vacuum energy is 
dependent on the cutoff $\Lambda$ via the $\mu^2\log(\Lambda/\mu)$ and 
$\log(\Lambda/\mu)$ terms in the string coupling expansion.  These two terms 
have a clear physical interpretation: $\log(\Lambda/\mu)$ is the effective 
volume of the Liouville dimension, and the tree-level and one-loop terms 
in the vacuum energy are proportional to this volume.  

The density of states in the M-theoy vacuum is similarly cutoff-dependent.   
The proper way of defining the double-scaling limit of $\rho$ involves first 
introducing a small-$\tau$ cutoff in the integral representation
\eqn\eedensdive{\rho_M(\mu)=\frac{1}{4\pi}\Re\int_{1/\Lambda}^\infty 
d\tau\,e^{-i\mu\tau}\frac{1}{\sinh^2(\omega_0\tau/2)},}
and then taking $\Lambda$ (which is proportional to $\sqrt{N}$) to infinity.  
In the previous subsections, we took advantage of the fact that the entire 
cutoff dependence of $\rho(\mu)$ is associated with the constant, 
$\mu$-independent term in $\rho(\mu)$, and we simply evaluated the finite, 
universal quantity $\p\rho/\p\mu$.  The leading, cutoff dependent term in 
$\rho$ is given by
\eqn\eedensdive{\frac{1}{4\pi\omega_0}\int_{1/\Lambda}^\infty 
d\tau\,\frac{1}{\sinh^2(\omega_0\tau/2)}=\frac{1}{2\pi\omega_0}\left(
\coth\left(\frac{\omega_0}{2\Lambda}\right)-1\right)\approx\frac{\Lambda}{\pi
\omega_0^2}+\ldots,}
where in the end we dropped all subleading terms in $1/\Lambda$.%
\foot{Throughout this paper, $\Lambda$ represents a large, nonuniversal 
cutoff. Consequently, we will only keep track of the leading dependence 
on $\Lambda$, and systematically drop all the subleading nonuniversal terms 
in all the cutoff-dependent quantities.}
This constant term modifies the leading behavior of the exact density of 
states in the $1/\mu$ expansion to 
\eqn\eerhocut{\rho_M(\mu)\approx\frac{1}{2\omega_0^2}(-\mu+\Lambda)+{\rm 
nonperturbative\ terms}.}
Upon further integration, the cutoff-dependent term in $\rho$ will give a 
$\Lambda$-dependent correction to our previous expression for the vacuum 
energy,
\eqn\eecutoffen{F\approx-\frac{\mu^3}{6\omega_0^2}
+\frac{\Lambda\mu^2}{4\omega_0^2}.}
In retrospect, we should have expected this $\Lambda$-dependent contribution 
to the leading behavior of the exact density of states, given the results of 
our WKB calculation in Section~5.1, where the same $\mu$-independent, 
$\Lambda$-dependent additive correction to $\rho\sim\mu$ was also found.    

In analogy with noncritical string theory, it is natural to interpret 
$\Lambda$ as a measure of the total volume of the system.  We see that in the 
noncritical M-theory vacuum -- just as in string theory -- the volume 
dependence creeps in via the $\mu^2$ term in the vacuum energy.  Unlike in 
string theory, however, this cutoff dependence does not affect the leading, 
tree-level term, which in M-theory scales as $\mu^3$.  The volume-dependent  
terms in the vacuum energy is now only subleading, of order $\kappa^{2/3}$ 
compared to the leading tree-level contribution.  It is intriguing to recall 
that in critical heterotic M-theory \hw, it is at this order where the twisted 
sector (described by Yang-Mills degrees of freedom at the boundary) starts 
contributing.  

\subsec{The Exact Formula}

Thus, the strong coupling expansion of $\rho_M$ in powers of $\mu$ results in 
a nontrivial series \eefirstsum .  This series can be summed as follows.  
Recall first that the Bernoulli numbers are usually defined via 
their generating function, 
\eqn\eeberndef{\frac{x}{e^x-1}=\sum_{n=0}^\infty\frac{B_nx^n}{n!}.}
Together with the elementary facts that $B_{2k+1}=0$ for all $k=1,2\ldots$ 
while $B_0=1$ and $B_1=-1/2$, this allows us to rewrite \eefirstsum\ as 
\eqn\eerhhighfin{\frac{\p\rho_M}{\p\nu}=-\frac{1}{\omega_0^2}\frac{\p}{\p\nu}
\left(\frac{\nu}{e^{2\pi\nu/\omega_0}-1}-\frac{B_1}{2\pi}2\pi\nu\right)=
-\frac{1}{\omega_0^2}\frac{\p}{\p\nu}\left(
\frac{\nu}{2}+\frac{\nu}{e^{2\pi\nu/\omega_0}-1}\right).}
To further verify this, we now evaluate \eederden\ directly, using a contour 
integral method while keeping track of the expected asymptotics in $\mu$.  

\subsubsec{Evaluation by a contour integral}

We are interested in
\eqn\eeinterested{I\equiv\frac{1}{4\pi}\int_0^\infty d x\,
\frac{e^{-i\mu x}}{\sinh^2(x/2)}\equiv\int_0^\infty d x\,I(x).}
This integral can be evaluated as follows.  The integrand $I(z)$, as a 
function in the complex plane, has an infinite series of double poles at 
$z=2k\pi i$ for all $k\in\Z$.  Furthermore, $I(z)$ is a quasi-periodic 
function along the imaginary axis,  
\eqn\eequasip{I(x+2\pi i)=e^{2\pi\mu}I(x).}
Taking advantage of this quasi-periodicity of $I(z)$, we can close the contour 
as in Fig~1,%
\fig{The integration contour $C$ used to evaluate the exact density of 
states.  The singularities are at $2k\pi i$, the contour encloses one of 
them -- at $2\pi i$ -- and the integral is evaluated in the limit of 
$L\rightarrow\infty$.}{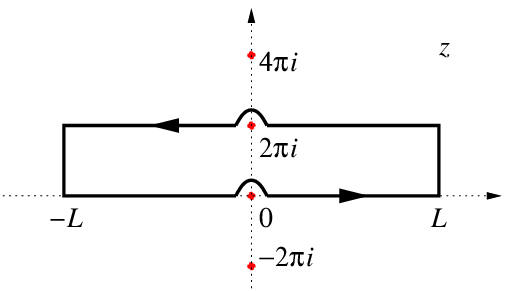}{3truein}
\noindent 
and obtain in the limit $L\rightarrow\infty$
\eqn\eeifirst{2(1-e^{2\pi\mu})I+\int_{C_0}I(z)\,dz+\int_{C_{2\pi i}}I(z)\,dz=
\oint_C I(z)=2\pi i\Res_{2\pi i}I(z)=2\mu e^{2\pi\mu}.}
where $C_0$ and $C_{2\pi i}$ are the two semicircles of radius 
$\epsilon\sim 1/\Lambda$ around the poles at 0 and $2\pi i$.  The 
contributions from $C_0$ and $C_{2\pi i}$ are divergent, and equal to 
\eqn\eesemicircles{\eqalign{\int_{C_0}I(z)&=-\frac{2}{\pi\epsilon}
-\pi i\Res_0I(z)+\CO(\epsilon),\cr
\int_{C_{2\pi i}}I(z)&=\frac{2}{\pi\epsilon}e^{2\pi\mu}
+\pi i\Res_{2\pi i}I(z)+
\CO(\epsilon).\cr}}
This yields
\eqn\eeifirsthalf{2(1-e^{2\pi\mu})I=\pi i\Res_{0}I(z)+
\pi i\Res_{2\pi i}I(z)+(1-e^{2\pi\mu})\frac{2}{\pi\epsilon},}
and finally (after identifying $\epsilon\sim1/\Lambda$)
\eqn\eeisecond{I=-\frac{\mu}{2\tanh(\pi\mu)}+\frac{\Lambda}{2}.}
Restoring $\omega_0$ in \eeisecond\ we get
\eqn\eerewri{\rho_M(\mu)=-\frac{\mu/\omega_0^2}{2\tanh(\pi\mu/\omega_0)}+
\frac{\Lambda}{2\omega_0^2}.}
This formula is exact, and matches the results of our summation 
\eerhhighfin .  One can easily check that it has the correct asymptotics 
to match both the weak coupling and the strong coupling expansion.  Notice 
that the exact density of states \eerewri\ is now even under 
$\mu\rightarrow-\mu$, despite the fact that the leading behavior in the 
asymptotic expansion in $1/\mu$ is odd, $\sim\mu^3$. This is the promised 
resolution of the puzzle mentioned in Section~5.1.  We see that the 
leading perturbative term being odd in $\mu$ is an artifact of splitting 
the exact formula into the perturbative and the nonperturbative part, neither 
of which are separately even under $\mu\rightarrow-\mu$.  

The exact vacuum energy is then 
\eqn\eepolylex{\eqalign{F&=-\frac{1}{2\omega_0^2}\int^{-\mu}
d\nu\left\{\frac{\nu^2}{\tanh(\pi\nu)}-\Lambda\omega_0\nu\right\}
=-\frac{1}{6\omega_0^2}\,\mu^3+\frac{\Lambda}{4\omega_0}\,\mu^2\cr
&\quad{}-\frac{1}{2\pi\omega_0}\,\mu^2\log(1-e^{-2\pi\mu/\omega_0})
+\frac{1}{2\pi^2}\,\mu\Li_2(e^{-2\pi\mu/\omega_0})
+\frac{\omega_0}{4\pi^3}\Li_3(e^{-2\pi\mu/\omega_0}).\cr}}
In the last expression, the first two terms correspond to the perturbative  
contribution in the $1/\mu$ expansion, while the terms involving the log and 
the polylogarithms%
\foot{Our notation for the polylogarithms is such that $\Li_\nu(z)=
\sum_{k=1}^\infty z^k/k^\nu$ for $|z|<1$.}
all represent nonperturbative corrections.  

In Section~5.3 we determined that the M-theory analog of the string 
susceptibility in the M-theory vacuum is equal to $1/2$.  This critical 
exponent measures the leading scaling of the vacuum energy with $\Delta$, 
a scaling variable that in string theory can be identified with the worldsheet 
cosmological constant.  Indeed, $\Delta$ is the continuum limit of the bare 
cosmological constant, which counts the number of vertices in the 
triangulation of the random surface in the matrix model.  $\Delta$ in the 
M-theory vacuum can also be exactly evaluated, leading to 
\eqn\eedeltamex{\Delta(\mu)=-\frac{\pi}{2}\left[\frac{1}{\omega_0^2}\mu^2
+\frac{2}{\omega_0}\mu\log(1-e^{-2\pi\mu/\omega_0})-
\Li_2(e^{-2\pi\mu/\omega_0})\right]+\frac{\pi\Lambda}{2\omega_0^2}\mu.}
We have again dropped a possible nonuniversal $\Lambda$-dependent constant 
term independent of $\mu$.  

\subsec{The Exact Formula in the Weak Coupling Expansion}

The compact formula \eepolylex\ for the exact vacuum energy can be rewritten 
in a more illuminating way by re-expanding in $1/\mu$, into an infinite 
sum of instanton-like terms, 
\eqn\eereexp{F=\omega_0\left\{-\frac{1}{6\omega_0^3}\,\mu^3+\frac{\Lambda}{4
\omega_0^2}\mu^2+\sum_{k=1}^\infty\left(\frac{1}{2\pi\omega_0^2k}\,\mu^2+
\frac{1}{2\pi^2\omega_0k^2}\,\mu+\frac{1}{4\pi^3 k^3}\right)
e^{-2\pi k\mu/\omega_0}\right\}.}
This formula exhibits several noteworthy features:

\item{$\bullet$} We again find that the weak-coupling expansion is in 
fractional powers of the natural loop counting parameter $\kappa^2$, the 
basic unit of the expansion being $\kappa^{2/3}$.

\item{$\bullet$} The strength of all nonperturbative effects (\ie , the 
``instanton action'') is controlled by $1/\mu\equiv\kappa^{2/3}$, leading to 
their scaling as ${}\sim e^{-A\kappa^{-2/3}}$ for some constants $A$.  

\item{$\bullet$} In the vicinity of each instanton, the perturbative expansion 
involves terms of order $\kappa^{2/3}$, $\kappa^{4/3}$, and $\kappa^0$ 
(which corresponds to one loop).  As in the case of the perturbative 
part of the vacuum energy, all higher orders terms vanish.  This is again 
strongly indicative of localization phenomena and an underlying topological 
symmetry of the theory.  It is intriguing, however, that this topological 
nature of the theory is compatible with the anticipated presence of a 
propagating degree of freedom: The M-theory analog of the massless stringy 
tachyon.%
\foot{A simple heuristic argument for the presence of a propagating degree of 
freedom can be given.  In the fermion language, the M-theory vacuum has a 
smooth semiclassical Fermi surface, of codimension one in phase space. This 
Fermi surface can fluctuate and its small fluctuations correspond to gapless 
bosonic excitations, implying the presence of a propagating degree of freedom.}
In this respect, the vacuum energy of the M-theory vacuum exhibits features 
reminiscent of a holographic field theory \refs{\hft,\phdm}.  

\item{$\bullet$} Unlike for the one-loop-exact perturbative contribution, 
the instanton measure contributions start only at the subleading order 
$\kappa^{2/3}$ compared to the tree-level term.  The analogy with critical 
heterotic M-theory \hw\ suggests that the instantons could be related to 
twisted sectors of the theory, with their characteristic subleading behavior 
$\sim\kappa^{2/3}$.  

\item{$\bullet$} Instead of calculating the vacuum energy $F$, one may be 
interested in the free energy $\Gamma(\mu)$, defined via the Legendre tranform 
of $F$ as a function of $\Delta$: 
\eqn\eelegendref{\Gamma(\mu)=\mu\Delta-F(\Delta).}
The exact formula for $\Gamma(\mu)$ can be easily obtained from \eedeltamex\ 
and \eereexp\ .  

\newsec{Analogy with the Debye Model of Phonons in Solids}

The universal part of our exact formula \eepolylex\ for the vacuum energy in 
the M-theory vacuum can be rewritten as 
\eqn\eevenmex{F=-\frac{1}{2\omega_0^2}\int^{-\mu}\nu^2\left(
\frac{1}{\exp\left(2\pi\nu/\omega_0\right)-1}+\frac{1}{2}\right)
d\nu.}
This expression is strongly suggestive of underlying bosonic degrees of 
freedom.  At first, it appears that \eevenmex\ represents the energy of a 
thermal bosonic system, with the density of states yielding Planck's black 
body radiation formula at an effective temperature of order one in string 
units,
\eqn\eetempeff{T_{\rm eff}=\frac{\omega_0}{2\pi}.}
However, \eevenmex\ also exhibits an effective cuttoff on the available 
frequencies, which is absent in the Planck black body formula.  Upon closer 
inspection, \eevenmex\ turns out to be more closely analogous to another 
famous bosonic system: The Debye model of phonon excitations in a solid at 
temperature \eetempeff . 

\subsec{The Debye Model} 

The Debye model of the thermodynamic properties of solids 
\refs{\debye,\lebellac} was originally designed to explain the behavior of the 
specific heat of solids at low temperatures.  It was 
proposed as an improvement of the somewhat less successful Einstein model, 
in which the atoms in the solid were simply treated as independent harmonic 
oscillators.  In Debye's model, the crystal consists of a fixed number 
$\sim\CN$ of atoms, assumed to behave as a system of coupled harmonic 
oscillators, with a fixed number $\CN$ of normal modes of the phonon 
spectrum.  The total energy of the system at temperature $T$ is given by 
an integral over all frequencies,  
\eqn\eedebyerev{E=\int_0^{\omega_D}\omega\,\rho_D(\omega)\left(
\frac{1}{e^{\omega/T}-1}+\frac{1}{2}\right)d\omega,}
with $\rho_D(\omega)$ the density of states of the system.  The finiteness of 
the total number of atoms imposes a limit $\omega_D$ on the maximum attainable 
frequency of the normal modes.  This limiting frequency, referred to as the 
Debye frequency, is set by the total number $\CN$ of normal modes in the 
crystal, 
\eqn\eedebyefreq{\int_0^{\omega_D}\rho_D(\omega)\,d\omega=\CN.}
As a model of realistic crystals, the Debye model is based on several rather 
drastic simplifying assumptions about the system.  Firstly, the phonon 
dispersion relation is assumed isotropic and strictly relativistic.  
Furthermore, the assumed absence of anharmonic terms in the phonon system 
is equivalent to ignoring the possibility of crystal melting at high enough 
temperature.  Lastly, the density of states $\rho_D(\omega)$ is assumed to 
be a smooth function $\rho_D(\omega)\sim\omega^{d-1}$, with $d$ the number 
of spatial dimensions, up to the sharp cutoff at $\omega=\omega_D$.  
For example, in two spatial dimensions, one would have
\eqn\eedebyedens{\rho_D(\omega)=\frac{V}{2\pi}\,\omega,}
 where $V$ is the volume of the system, and the speed of sound has been set 
equal to one.  

All of these assumptions would have to be modified in a realistic crystal.  
In contrast, as we are now going to see, the exact calculations in M-theory 
are compatible with all of the above assumptions, and in this sense, the Debye 
analogy for noncritical M-theory is exact.   

\subsec{The Analogy}

It is easy to see that our formula \eevenmex\ for the exact vacuum energy of 
M-theory is precisely of the Debye form \eedebyerev , with the following 
dictionary between the two descriptions of the system:

\item{$\bullet$} 
The chemical potential in the bosonic system is equal to zero.  This means 
that the bosonic quanta can be created and annihilated, and that the total 
number of bosons is not fixed.  

\item{$\bullet$} The perturbative piece in $\rho(\mu)$ corresponds to the 
zero-point energy of the Debye crystal.  The nonperturbative terms in 
$\rho(\mu)$ sum up to the Planck thermal factor.  

\item{$\bullet$} The double-scaled Fermi energy $\mu$ plays the role of the 
Debye frequency $\omega_D$. 

\item{$\bullet$} The effective temperature of the crystal is given by 
\eetempeff.  It is set in string units and cannot be varied, at least in this 
vacuum of noncritical M-theory.  

\item{$\bullet$} The Debye density of states is proportional to $\omega$.   
Thus, the system is effectively $2+1$ dimensional, as suggested by its 
M-theory interpretation as the M-theory vacuum.  

\item{$\bullet$} The formula is consistent with the exact relativistic 
dispersion relation of the phonons, and with the relativistic density of 
states $\rho_D(\omega)=V\omega/2\pi$.  

\item{$\bullet$} The total volume $V$ of the Debye crystal is proportional to 
$1/\omega_0^2$.  However, the surprise lies in the overall sign of this 
volume, which comes out negative!

This last point can be better understood as follows.  Note first that the 
effective Debye density of states $\rho_D(\omega)$ that appears in 
\eedebyerev\ can be identified with the leading perturbative term in our 
density of states $\rho(\mu)$:
\eqn\eeleadpertd{\rho(\omega)\sim\rho_D(\omega)+{\rm nonperturbative\ terms}.}
Recalling now the relation between the scaling variable $\Delta$ and the 
density of states, 
\eqn\eedeldenag{\Delta=\pi\int^{-\mu}\rho(\nu)\, d\nu,}
 the leading perturbative term in $\Delta$ is found to be related, via 
\eedebyefreq , to the total number of atoms in the Debye solid:
\eqn\eedeltasolid{\Delta\sim \CN+{\rm nonperturbative\ terms}.}
In string theory, the scaling variable $\Delta$ was interpreted as 
the worldsheet cosmological constant, since its discretized matrix-model 
version counts the number of plaquettes in the random triangulation of the 
worldsheet.  Surprisingly, we see that even in noncritical M-theory, 
$\Delta$ can be interpreted as an object that counts the number of 
constituents, now of the Debye crystal.  

\subsec{Reintroducing the Cutoff: The Melting Crystal Interpretation}

The Debye analogy is almost precise, except that -- as we have just seen -- 
it seems to lead to the rather embarrassing prediction of a negative volume 
for the Debye crystal.  This problem can be remedied by reintroducing the 
dependence on the cutoff $\Lambda$ in the system.  

We have seen in our exact evaluation of $\Delta$ in Section~6.4 
that when we keep track of the cutoff dependence, $\Delta$ 
gets a large positive contribution proportional to $\Lambda\mu$,
\eqn\eedeltamex{\Delta(\mu)=-\frac{\pi}{2\omega_0^2}\mu^2
+\frac{\pi\Lambda}{2\omega_0^2}\mu.}
As we have just argued, in the Debye model analogy, $\Delta$ counts the 
effective number of atoms in the Debye crystal.  Hence, \eedeltamex\ 
shows that in the thermodynamic limit of large $\Lambda$, we effectively 
have a a large Debye crystal whose number of atoms is measured by the cutoff 
$\Lambda$.  The negative sign in front of the leading $\mu^2$ term in $\Delta$ 
is now easily understood:  A number of atoms, measured by $\mu$, has 
been removed from the large crystal.%
\foot{In fact, this is closely analogous to the behavior of noncritical 
string theory, where the available volume of the Liouville dimension is 
measured by $\log(\Lambda/\mu)$. One can think of $\Lambda$ as setting the 
size of the Liouville dimension in the weakly coupled asymptotic region.  
$\mu$ is then associated with the Liouville wall.  At weak string coupling 
$\mu\gg 1$, Liouville wall effectively subtracts the available volume from 
the total volume set by $\Lambda$, similarly to the behavior we have observed 
in noncritical M-theory.}
$\mu$ now represents the {\it lowest\/} frequency in the system, confirming 
that a small number of atoms has been {\it removed\/} from the Debye solid.  
Effectively, $\mu$ measures the size of a small hole in a big sample of the 
Debye crystal.  This picture is superficially reminiscent of the recently 
found correspondence between topological strings and the statistical mechanics 
of a classical melting crystal \melting .

Having interpreted the cutoff $\Lambda$ as the quantity that sets the total 
number of atoms in the Debye system, we can in fact sharpen the relation 
between $\Delta$, $\CN$ and $N$ even further.  The cutoff-dependent terms in 
$\Delta$ will include a $\mu$-independent constant, which we have been 
ignoring so far as nonuniversal.  Restoring this term, we get
\eqn\eedeltaleadlam{\Delta\sim \frac{\pi}{2\omega_0^2}(-\mu^2+\Lambda^2)+
\ldots.}
The $\Lambda^2$ term can be thought of as coming from the lower integration 
bound in the definition of $\Delta$ in \eedeldenag .  Recalling now that in 
the large $N$ limit, the nonuniversal cutoff $\Lambda$ scales as $\sqrt{N}$, 
we obtain from \eedeltaleadlam\ that
\eqn\eecnen{\CN\sim N.}
Thus, we find that the number of atoms in the Debye solid is effectively 
related to the number of fermions in the Fermi liquid.  

\subsec{Solution with Two Fermi Surfaces as a Universal Melting Crystal}

The reinterpretation of $\mu$ as a parameter measuring the number of atoms 
removed from a Debye crystal of size set by the cutoff $\Lambda$ is pleasing, 
but the downside of this interpretation is in its reliance on the nonuniversal 
cutoff $\Lambda$.  In particular, it would be desirable to have a more 
detailed information about the bulk of the system.  For example, we would like 
to know whether the large crystal is at the same temperature as the atoms 
removed from it.    

The dependence of the Debye interpretation on the cutoff can be eliminated 
by considering a small modification of our construction of the M-theory vacuum 
state.  Instead of using the nonuniversal cutoff $\Lambda$ to provide the 
environment, introduce two Fermi levels, $\mu_\pm$, with $\mu_+<\mu_-$, and 
fill the Fermi sea only between $\mu_+$ and $\mu_-$.  Hence, $\mu_+$ and 
$\mu_-$ are the top and the bottom of the Fermi sea, respectively.  This 
state is again interpreted in terms of the double-scaling limit:  Define 
$\mu_\pm=-N\varepsilon_F^\pm$, and take the limit $N\rightarrow\infty$ and 
$\varepsilon_F^\pm\rightarrow 0$ while keeping $\mu_\pm$ fixed.    

In this modified state $|\,{\rm M},\mu_+,\mu_-\rangle$, the universal part of 
the vacuum energy is 
\eqn\eeexpm{F=-\frac{2}{\omega_0^2}\int_{-\mu_-}^{-\mu_+}d\nu\,
\nu^2\left(\frac{1}{\exp\left(2\pi\nu/\omega_0\right)-1}+\frac{1}{2}
\right),}
while $\Delta$ is
\eqn\eedeltapm{\Delta(\mu_+,\mu_-)=\frac{\pi}{2\omega_0^2}(\mu_-^2-\mu_+^2)+
\ldots,}
where the ``$\dots$'' stand for all the nonperturbative and nonuniversal 
terms.  

\eeexpm\ is indeed the Debye result for the free energy of a crystal of size 
set by $\mu_-$, with a portion of the crystal measured by $\mu_+$ removed.  
If $\mu_-\gg\mu_+$, we have a small hole in a big Debye crystal.  The 
dependence of all quantities on $\mu_\pm$ is universal.  The system 
is at finite temperature of order one in string units, 
$T_{\rm eff}=\omega_0/2\pi$.  

It is natural to suspect that the bosonic features of the vacuum energy 
in the noncritical M-theory vacuum are related to the anticipated bosonization 
in terms of a collective degree of freedom, which should represent the 
M-theory lift of the massless tachyon of noncritical string theory.  
This connection, and the entire Debye analogy, deserves further study.  

\newsec{Observables and Symmetries}

Having discussed properties of a specific M-theory solution in the previous 
sections, we now address several more conceptual aspects of noncritical 
M-theory, which should find applications to a broader class of solutions.  
For the rest of the paper, we will set $\omega_0=1$.  

\subsec{Observables}

We have seen that the exact vacuum energy has a very interesting structure, 
suggesting an underlying symmetry reminiscent of topological locatization.  

Despite appearances, and the suggestive simplicity of the exact vacuum 
energy, the M-theory vacuum still contains propagating degrees of freedom.  
The existence of a Fermi surface suggests that at least one field-theory 
degree of freedom is present.  In string theory, the fluctcuations of the 
Fermi surface correspond to the massless modes of the theory, \ie , the 
tachyon (and, in Type 0B, also the RR scalar).  Motivated by how the 
tachyon emerges from the matrix models of two-dimensional string theories 
(see, \eg , \gmrev), we can define a set of natural observables given by the 
density of eigenvalues $\rho(t,\lambda_i)=\Psi^\dagger\Psi(t,\lambda_i)$ 
\djcoll , 
or, more conveniently, by the inverse Laplace-like transform of $\rho$ with 
respect to the eigenvalue coordinates, 
\eqn\eeobsnat{\CO_0(t,w_i)=\int d\lambda_1d\lambda_2\,
e^{-w_1\lambda_1-w_2\lambda_2}\Psi^\dagger\Psi(t,\lambda_i),}
or the Fourier transform with respect to $t$, 
\eqn\eeobsnat{\CO_0(\omega,w_i)=\int dt\, e^{i\omega t}\CO_0(t,w_i).}
Lessons learned in two-dimensional string theory lead us to anticipate that 
the field $\CO_0$ should be the M-theory analog of the massless tachyon 
field.  Indeed, it is this collective bosonic field that represents the 
fluctuations of the Fermi surface in circumstances where the latter is nicely 
defined.  
An even better representation of the observables is
\eqn\eeobsnato{\eqalign{\CO(t,\ell,\phi)&=\int_{\sqrt{2\mu}}^\infty 
d\lambda e^{-\ell\lambda}\Psi^\dagger\Psi(t,\lambda,\phi)\cr
&=\int_{0}^\infty d\lambda e^{-\sqrt{2\mu}\ell\cosh\tau }\Psi^\dagger\Psi
(t,\tau,\phi),\cr}}
where we have introduced $\tau$ via $\lambda=\sqrt{2\mu}\cosh\tau$ in order to 
shift the lower integration bound to zero.  These formulas are very 
reminiscent of the bosonization of nonrelativistic fermions in higher 
dimensions \refs{\haldane,\bosonization}, where the bosonization is in terms 
of a collection of $1+1$ dimensional bosons parametrized by the angle $\phi$ 
on the Fermi surface, which plays the role of an internal index.  

The natural correlation functions to calculate are the $n$-point functions
\eqn\eenpoint{\langle\,{\rm M},\mu|\,\prod_{k=1}^n\CO(t_k,\ell_k,
\phi_k)\,|\,{\rm M},\mu\rangle.}
They can again be evaluated exactly (in principle), using the techniques 
developed in the matrix models of noncritical strings 
\refs{\moore,\moorepr,\szerob}.  We leave a detailed analysis for the future.  

\subsec{Symmetries}

Our noncritical M-theory is an exactly solvable system, with an infinite 
dimensional symmetry algebra generalizing the famous $w_\infty$ symmetries 
and ground ring structure of two-dimensional strings \grring.  

Recalling the classical equations of motion
\eqn\eeclasseom{\dot p_i=\lambda_i,\qquad\dot\lambda_i=p_i,}
the conserved charges can be built out of four building blocks (interpreting 
$t$ again as a real time coordinate), 
\eqn\eeblocks{\eqalign{\sla_1&=\frac{1}{\sqrt{2}}(p_1+\lambda_1)e^{-t},\cr 
\sla_2&=\frac{1}{\sqrt{2}}(p_2+\lambda_2)e^{-t},\cr}\qquad
\eqalign{\slb_1&=\frac{1}{\sqrt{2}}(p_1-\lambda_1)e^{t},\cr 
\slb_2&=\frac{1}{\sqrt{2}}(p_2-\lambda_2)e^{t}.\cr}}
The full symmetry algebra $\CW$ is generated by products of non-negative 
integral powers of $\sla_i,\slb_i$.  Hence, a basis in $\CW$ is given by 
\eqn\eesymbasis{W_{m_1m_2n_1n_2}= \sla_1^{m_1}\sla_2^{m_2}\slb_1^{n_1}
\slb_2^{n_2},\qquad m_i,n_i=0,1,\ldots}
The commutation relations are defined via the elementary Poisson brackets, 
\eqn\eefish{[\sla_i,\slb_j]=-\delta_{ij},\qquad[\sla_i,\sla_j]=[\slb_i,\slb_j]
=0.}
Note that the Hamiltonian and the angular momentum are both bilinear 
combinations of $\sla_i$ and $\slb_i$:
\eqn\eequadlings{H=W_{1100}+W_{0011}\qquad J=W_{1001}-W_{0110}.}  

The elements in $\CW$ at most bilinear in $\sla_i$ and $\slb_i$ form a closed 
finite-dimensional subalgebra $\CW_0$ of the full infinite symmetry algebra 
$\CW$ of the system.  

In a typical solution, some symmetries from $\CW$ or $\CW_0$ respect the Fermi 
surface, while others are broken by the solution.  For example, our M-theory 
vacuum is preserved by just four (out of the total number of ten) quadratic 
charges: $W_{1010}, W_{1001}, W_{0110}$ and $W_{0101}$.  They form the algebra 
of $SO(3)\times U(1)$, with the Abelian generator corresponding to the 
Hamiltonian that defines the Fermi surface.  

\subsubsec{Massless modes vs.\ symmetries}

In a given solution of noncritical M-theory, the massless bosonic modes are 
closely related to the existence of the Fermi surface.  It is tempting to 
speculate that these massless modes should be interpreted as the Goldstone 
modes of the symmetries in $\CW$ that have been broken by the Fermi surface.  
Indeed, the bosonic fluctuations of the Fermi surface have been interpreted as 
Goldstone modes of broken symmetries in the condensed matter context (see, 
\eg , \haldane).  

If this view is correct, 
the states that exhibit higher degrees of symmetry should have fewer massless 
modes.  The M-theory vacuum indeed exhibits a larger symmetry than the 
Type 0A or 0B solutions.  This larger degree of symmetry could explain 
the apparent topological features of the exact vacuum energy, compared to the 
less symmetric Type 0A or 0B string vacua.  In this sense, 
the M-theory vacuum is closer than the string vacua to exposing the full 
symmetry of the theory.%
\foot{Note, however, that any nontrivial Fermi surface will 
always break at least {\it some\/} of the $\CW$ symmetry, and it is thus 
not clear whether the theory has a ground state in which the entire 
underlying symmetry is unbroken.}
This hypothetical Goldstone boson interpretation of the massless tachyon seems 
further supported by our interpretation of the vacuum energy in the 
M-theory vacuum in terms of the Debye phonons.  

\newsec{Semiclassical Spacetime Physics as Hydrodynamics of the Fermi Liquid}

As we have argued, only the solutions of noncritical M-theory that can 
be bosonized in terms of hydrodynamic degrees of freedom are expected to 
admit a conventional semiclassical spacetime description.  In this section, 
we develop a formalism -- closely parallel to a similar framework in 
noncritical string theory \refs{\polceeone,\djordje} -- which allows us to 
search systematically for such hydrodynamic solutions of noncritical 
M-theory.  Intuitively, the hydrodynamic states are those states that can be 
described by a semiclassical Fermi surface.  In the semiclassical limit, the 
Fermi surface satisfies its own hydrodynamical equations of motion.  Solving 
those equations directly is an efficient way of finding solutions of M-theory 
which admit a hydrodynamic description by design.  

\subsec{Classical Equations of Motion for the Fermi Surface}

The classical equations of motion for the Fermi surface in noncritical 
M-theory can be derived using the methods developed in noncritical string 
theory.  In the classical limit, the Fermi surface is a (possibly 
time-dependent) hypersurface in the four-dimensional phase space of the 
system. The location of the Fermi surface in phase space can be described, 
for example, by choosing $p_1$ as the dependent variable, 
\eqn\eepeeone{p_1\equiv P(x,y,w,t).}
Here we have relabeled $\lambda_1\equiv x$, $\lambda_2\equiv y$, and 
$p_2\equiv p_y\equiv w$, and have Wick-rotated $t$ back to real time.  
Repeating the steps used in noncritical string 
theory, one can show that this function $P$ satisfies the following classical 
equation of motion, 
\eqn\eefseom{\p_tP=x-P\p_xP-w\p_y P-y\p_wP.}
Sometimes it is convenient to use an alternative equation for the Fermi 
surface in the polar coordinates, in which the phase space is parametrized by 
$r,\phi$ and their canonically conjugate momenta $p_r$ and $p_\phi\equiv J$.  
As our dependent variable to describe the Fermi surface, we can choose 
$p_r\equiv\CP(r,\phi,p_\phi,t)$.  The equations of motion for $\CP$ are 
\eqn\eefseomp{\p_t\CP=r+\frac{p_\phi^2}{r^3}-\CP\p_r\CP-\frac{u}{r^2}
\p_\phi\CP.}
In the rest of this section, we will study several time-independent solutions 
of the theory, leaving time-dependent solutions for Section~10.  
Needless to say, our selection of solutions is just a small sampling.  

\subsec{Vacua with q as the Scaling Variable}

The conventional Fermi surface that defined our M-theory vacuum in much of 
this paper,
\eqn\eeconvfs{\frac{p_r^2}{2}+\frac{p_\phi^2}{2r^2}-\frac{r^2}{2}=-\mu}
satisfies the equation of motion, with
\eqn\eepconvfs{\CP(r,\phi,u,t)=\sqrt{r^2-\frac{p_\phi^2}{r^2}-2\mu}.}

So does a Fermi surface given by filling up to a fixed value of another 
conserved quantity, the angular momentum:
\eqn\eefsq{p_\phi=q.}
However, the fluctuations around this surface are inconveniently parametrized 
in our representation of the Fermi surface by $\CP$.  We revert to the 
Cartesian coordinates, where this same surface is parametrized by
\eqn\eefsqc{p_1\lambda_2-p_2\lambda_1=q,}
leading to 
\eqn\eepfsqc{P(x,y,w,t)=\frac{q+wx}{y}.}
This satisfies the classical equation of motion for the Fermi surface in 
the Cartesian coordinate representation.  We expect this solution to be 
related to two-dimensional string backgrounds with $q$ as the scaling variable 
\refs{\deformed,\danielsson,\kapustin,\gtato} or to $AdS_2$ backgrounds 
\refs{\hermanads,\andyads}.  

\subsubsec{Duality to Thermofield Dynamics in the Rightside-Up Harmonic 
Potential}

It turns out that a simple canonical transformation of the variables of our 
model maps our system to the thermofield dynamics of second-quantized fermions 
in the rightside-up harmonic potential.  

The Fermi surface that fills all sectors up to a fixed $q$ can be rewritten 
as follows.  Define
\eqn\eenewvar{\eqalign{x'&=\frac{1}{\sqrt{2}}(x+p_y),
\qquad p_{x'}=\frac{1}{\sqrt{2}}(p_x-y),\cr
y'&=\frac{1}{\sqrt{2}}(y+p_x),\qquad p_{y'}=
\frac{1}{\sqrt{2}}(p_y-x).\cr}}
In these new variables, the Fermi surface is
\eqn\eenewvarfs{\frac{1}{2}\left[p_{x'}^2+(x')^2-p_{y'}^2-(y')^2\right]=-q.}
This is simply a system consisting of two regular rightside-up harmonic 
oscillators, with a relative sign between the two Hamiltonians.  Such a 
combination of two copies of the same Hamiltonian with a relative minus 
sign defines the real-time thermofield dynamics of the system (see, 
\eg , \refs{\schw-\veronika} for some background).  Hence, we find 
the rather surprising result, that our noncritical M-theory is dual to the 
thermofield dynamics of double-scaled fermions in the rightside-up 
harmonic oscillator potential.   

A double-scaling limit of $1+1$ dimensional fermions in the rightside-up 
harmonic potential has been studied, as the nonperturbative definition of 
a somewhat exotic version of $c=1$ string theory \refs{\itzmcg,\boyar}, 
(see also \refs{\cjr,\berenstein} for another possible viewpoint).  
Here we see that this string theory is naturally embedded into our framework 
of noncritical M-theory.  Indeed, the string theories of \refs{\itzmcg,\boyar} 
can be obtained as 
solutions by repeating the steps of Section~4 in the primed variables.  For 
example, filling only the states with a fixed value of $\nu_2'$ will produce 
the string theory of the rightside-up harmonic oscillator studied in 
\refs{\itzmcg,\boyar}.  

Thermofield dynamics of a given system is not defined just by the doubling 
of the degrees of freedom and specifying the Hamiltonian.  An important part 
of the definition is the preparation of an entangled vacuum state.  In our 
case, this thermal state is
\eqn\eethfd{\sum e^{-E_\Phi/T}|\Phi\rangle\otimes|\widetilde\Phi\rangle,}
where the sum is performed over all quantum states $|\Phi\rangle$ of the 
second-quantized rightside-up harmonic oscillator, with $E_\Phi$ the energy 
of $|\Phi\rangle$).  In accord with the philosophy of Section~3.3, this 
thermal state of the thermofield dynamics of the rightside-up harmonic 
oscillator will be on the moduli space of all solutions of noncritical 
M-theory.  

We also note in passing that if one performs the particle-hole duality 
on just one of the two upside-down oscillators that define noncritical 
M-theory, the Hamiltonian becomes that of the thermofield dynamics of one 
upside-down harmonic oscillator.  

\subsec{A Family of Stationary Solutions}

Clearly, a bigger class of time-independent classical solutions is obtained by 
combining the two conserved quantities, $E$ and $J$, and postulating a 
Fermi surface
\eqn\eefsclass{\frac{1}{2}\left(p_x^2+p_y^2-x^2-y^2\right)
+\Omega\left(p_xy-p_yx\right)=-\mu,}
where $\Omega$ is a constant parameter.  Since $\Omega$ serves as the 
chemical potential for the conserved angular momentum $J$, it can be 
interpreted as the angular velocity, leading to a simple interpretation of 
this solution as uniformly rotating.  The vacuum energy of this state can 
again be evaluated exactly, as follows.  In the polar coordinate 
representation of the model, 
the Fermi surface \eefsclass\ can be viewed in each sector of fixed 
$J=q$ as the Type 0A theory with RR flux $q$ and the Fermi sea filled up to 
a $q$-dependent Fermi level, effectively replacing $\mu$ by $\mu+\Omega q$.  
The summation over all values of $q$ then leads to 
\eqn\eedensmomega{\eqalign{\frac{\p\rho(\mu,\Omega)}{\p\mu}&=
\frac{1}{2\pi\omega_0}\Im\int_0^\infty d\tau\sum_{q\in\Z}e^{-i(\mu
+\Omega q)\tau}\frac{\omega_0\tau}{\sinh(\omega_0\tau)}\,e^{-|q|\omega_0\tau}
\cr
&=\frac{1}{2\pi\omega_0}\Im\int_0^\infty d\tau\,e^{-i\mu\tau}
\frac{\omega_0\tau}{\cosh(\omega_0\tau)-\cos(\Omega\tau)},\cr}}
where we have temporarily restored the dependence on $\omega_0$.  
Integrating \eedensmomega\ once, we get
\eqn\eedensmom{\rho(\mu,\Omega)=\frac{1}{2\pi}\Re\int_0^\infty 
d\tau\,e^{-i\mu\tau}\frac{1}{\cosh(\omega_0\tau)-\cos(\Omega\tau)}.}
This again requires a cutoff at the lower integration limit $\tau\sim 0$.   

In this family of solutions, the two conserved time-independent charges $H$ 
and $J$ have been essentially put on an equal footing.  The main difference 
between them is that one of them is compact and the other one is not.  
In string theory, the $\mu$ and $q$ are related to the string coupling and 
the RR flux, respectively, but from the higher-dimensional vantage point of 
noncritical M-theory they are much more closely related.  We believe that this 
M-theory perspective may be at the core
of some of the surprising patterns observed recently in the behavior of 
two-dimensional strings in \juannati .

\subsec{A Twisted M-Theory State}

There is a simple variation of the M-theory state, which illustrates several 
interesting points.  This state $|\,\widetilde{\rm M},\mu\rangle$ is defined 
by filling all states in sectors with even angular momentum $q$ up to a Fermi 
surface $-\mu$, while filling all sectors with odd $q$ {\it down\/} to $-\mu$:
\eqn\eetwistedm{\eqalign{a_q(\nu)\,|\,\widetilde{\rm M},\mu\rangle&=0\qquad
{\rm for}\ \left\{\eqalign{\nu&>-\mu,\quad q\ {\rm even},\cr
\nu&<-\mu,\quad q\ {\rm odd},\cr}\right.\cr
a_q^\dagger(\nu)\,|\,\widetilde{\rm M},\mu\rangle&=0\qquad{\rm for}\ \left\{
\eqalign{\nu&<-\mu,\quad q\ {\rm even},\cr
\nu&>-\mu,\quad q\ {\rm odd}.\cr}\right.\cr}}
The calculation of the vacuum energy goes through as in the case of 
$|\,{\rm M},\mu\rangle$, with an additional $(-1)^q$ weighing the contribution 
of each sector of fixed $q$.  This will change the density of states to
\eqn\eemoddens{\rho_{\widetilde M}=\frac{1}{4\pi}\Re\int_0^\infty d\tau\,
e^{-i\mu\tau}\frac{1}{\cosh^2{(\omega_0\tau/2)}}.}
This integral can again be evaluated exactly,%
\foot{The integration contour is again that of Fig.~1, but the poles of the 
integrand are now at $\pi(2k+1)i$ for $k\in\Z$.  Hence, the radii of the 
two semi-circles can be taken to zero, and only the pole at $\pi i$ 
contributes to the integral.}
leading to 
\eqn\eetwistedrho{\rho_{\widetilde M}=\frac{1}{2\pi\omega_0^2}\,
\frac{\mu}{\sinh(\pi\mu/\omega_0)}.}
This solution exhibits some interesting points: 

\item{$\bullet$} Unlike in the case of the M-theory state 
$|\,{\rm M},\mu\rangle$, the density of states and the vacuum energy of 
the twisted M-theory state are cutoff independent.  Moreover, the leading 
term $\sim\mu$ in the $1/\mu$ expansion of the density of states is absent.  

\item{$\bullet$} In fact, the expression is fully nonperturbative in the 
$1/\mu$ expansion.  The exact vacuum energy consists of an infinite series 
of nonperturbative terms,
\eqn\eetwistedvacen{F=-\frac{1}{\pi^4\omega_0^2}\sum_{k=0}^\infty 
\left(\frac{1}{2k+1}\,\left(\frac{\pi\mu}{\omega_0}\right)^2
+\frac{2}{(2k+1)^2}\,\frac{\pi\mu}{\omega_0}+\frac{2}{(2k+1)^3}
\right)e^{-(2k+1)\pi\mu/\omega_0}.}
It would be desirable to identify the precise symmetry (perhaps akin to 
supersymmetry) responsible for the exact vanishing of the vacuum 
energy to all orders in perturbation theory, but perhaps violated by the 
nonperturbative effects.  

\item{$\bullet$} $|\,\widetilde{\rm M},\mu\rangle$ should clearly be 
considered a hydrodynamic state, although it is somewhat outside of the class 
of hydrodynamic states that solve the equations of motion for the 
semiclassical Fermi surface.  Indeed, due to the staggered manner of how 
states of different $q$ are filled, the average density of fermions is 
continuous across the surface of Fermi energy $-\mu$.  Perhaps a more useful 
semiclassical observable would be the staggered density of eigenvalues, 
$\widetilde\CO$, defined as in \eeobsnato\ with an additional insertion of 
$(-1)^q$ in each sector of angular momentum $q$.  

\newsec{Time-Dependent Solutions}

We can generate some time-dependent solutions by continuing the strategy 
from the previous section.  In particular, we can modify a given Fermi 
surface by adding conserved quantities that explicitly contain $t$.  This 
is very similar to the strategy used in noncritical string theory in 
\refs{\karczs-\das}.  We can immediately write an family of time-dependent 
solutions, by simply postulating a Fermi surface
\eqn\eegentime{\sum_{n_i,m_i=0}^{\infty}\tau_{m_1m_2n_1n_2}W_{m_1m_2n_1n_2}
(p_i,\lambda_i,t)=0,}
where $W_{m_1m_2n_1n_2}$ is the basis \eesymbasis\ of the $\CW$ symmetry 
algebra, and $\tau_{m_1m_2n_1n_2}$ are arbitrary constants.  Note that 
$\tau_{0000}$ effectively plays the role of the scaling variable $\mu$, 
since it multiplies the central element $W_{0000}\sim 1$ of the symmetry 
algebra.  

The family of static solutions \eefsclass\ is in this class, with only 
$\tau_{1100}=-\tau_{0011}$, $\tau_{1001}=\tau_{0110}$, and $\tau_{0000}$ 
nonzero.  

We can now look at some examples of time-dependent solutions from this class.  

\subsec{Losing or Gaining a Dimension}

The simplest time-dependent solutions are obtained by adding to the 
Hamiltonian terms linear in $\sla_i$ and $\slb_i$. Of such solutions, the 
simplest will give the following time-dependent Fermi surface,
\eqn\eetimed{\frac{1}{2}\left(p_x^2+p_y^2-x^2-y^2\right)+c(p_y-y)e^t=-\mu,}
where $c=\tau_{0001}/\sqrt{2}$ is a constant.  In the asymptotic past, 
$t\rightarrow-\infty$, the effect of the time-dependent deformation is 
negligible, and the Fermi surface approaches the static Fermi surface of the 
M-theory vacuum in $2+1$ dimensions.  At late times $t\rightarrow\infty$, 
however, the Fermi sea is partially drained.  Another, similar solution is 
given by
\eqn\eetimedd{\frac{1}{2}\left(p_x^2+p_y^2-x^2-y^2\right)+\frac{\tilde c}{2}
(p_y-y)^2e^{2t}=-\mu,}
wit $\tilde c=\tau_{0002}$ again a constant.  This again describes a solution 
that starts off as the M-theory vacuum, whose Fermi sea is drained at late 
times everywhere except along the hypersurface $y=p_y$, where the Fermi sea 
stays at $-\mu$.  Along this hypersurface, the conserved quantity 
$\nu_y\equiv\sla_2\slb_2$ vanishes.  Recalling our construction of the Type 0B 
string theory vacuum in M-theory, in which only states with $\nu_y=0$ were 
filled up to Fermi level $-\mu$, it is natural to identify the time-dependent 
solution \eetimedd\ as decaying at late times into the Type 0B vacuum.  
In the process, the effective spacetime dimension changes from $2+1$ to 
$1+1$.  

Similarly, solutions with $\tau_{0010}$ or $\tau_{0020}$ nonzero will 
correspond to the time reversal of \eetimed\ and \eetimedd ,   
\eqn\eetimedr{\frac{1}{2}\left(p_x^2+p_y^2-x^2-y^2\right)-c(p_y+y)e^{-t}=-\mu,}
and
\eqn\eetimeddr{\frac{1}{2}\left(p_x^2+p_y^2-x^2-y^2\right)
+\frac{\tilde c}{2}(p_y+y)^2e^{-2t}=-\mu.}
In particular, \eetimeddr\ can be interpreted as the time-dependent Fermi 
surface of a solution that starts off as the Type 0B vacuum at early times, 
and then evolves into the M-theory vacuum at late times.  

\subsec{Solutions Interpolating Between Two String Vacua}

The ingredients of time-dependent solutions from the previous subsection can 
be easily combined, to construct a solution interpolating between two string 
theories, via an intermediate M-theory phase.  Consider for example
\eqn\eefulltime{\eqalign{\frac{1}{2}\left(p_x^2+p_y^2-x^2-y^2\right)
&+\frac{1}{2}\left[c_1(p_x-x)^2+c_2(p_y-y)^2\right]e^{2t}\cr
&+\frac{1}{2}\left[c_3(p_x+x)^2+c_4(p_y+y)^2\right]e^{-2t}=-\mu.\cr}}
Here $c_1,\ldots, c_4$ are again constants that can be chosen arbitrarily.   
With only $c_1$ and $c_4$ nonzero and positive, \eefulltime\ is the Fermi 
surface of a time-dependent solution that starts at early times as Type 0B 
with $x$ playing the role of the spatial dimension, and decays at late times 
into another Type 0B vacuum, now with $y$ playing the role of the spatial 
dimension.  At times of order $t\approx 0$, this solution is going through a 
$2+1$ dimensional M-theory phase, with the Fermi surface filled more 
democratically in the $x,y$ plane.  

In principle, even though the spacetime dimension may be changing, 
the free fermion formulation still defines a unitary quantum evolution, 
and can be used to define an S-matrix between initial and final states, 
as defined in the asymptotic Type 0B string vacua where they are represented 
by the massless modes of the Type 0B tachyon.  This is a novelty compared to 
time-dependent solutions found in two-dimensional string theory 
\refs{\karczs-\das}: We can now have ``decays'' of spacetime with 
well-understood initial {\it and\/} final states simultaneously, both being 
described by a known semiclassical string vacuum.  Cosmological decays 
into ``nothing'' are also possible, for example with both $c_1$ and $c_2$ 
nonzero.  

Clearly, vast families of similar solutions exist, and one can engineer 
solutions that for example begin in the Type 0A vacuum and evolve into 
the Type 0B vacuum.  

\newsec{Conclusions}

In this paper, we have presented a fully nonperturbative definition of 
noncritical M-theory in $2+1$ dimensions, in terms of a double-scaling limit 
of a nonrelativistic Fermi liquid.  Clearly, in our analysis of this theory, 
we have only scratched the proverbial surface.  The exact solvability of the 
model allows one to extract a wealth of data about this incarnation of 
M-theory, including detailed information about some of its exotic 
time-dependent solutions.  

The theory is fully defined as a quantum-mechanical theory in terms of the 
fermions, even in regimes where a semiclassical spacetime interpretation 
ceases to be valid.  In this picture, the fundamental degrees of freedom 
of M-theory are the elementary fermions of the Fermi liquid.  

The fundamental fermions originate in the underlying system of D0- and 
anti D0-branes of the two-dimensional Type 0A string theory.  In this respect, 
our nonperturbative definition of noncritical M-theory bears striking 
resemblance to M(atrix) theory \refs{\bfss-\polmm} -- another candidate for 
a nonperturbative formulation of M-theory, defined in terms of the 
supersymmetric quantum mechanics of $N$ D0-branes of Type IIA string theory 
in the Sen-Seiberg scaling limit.  A possible relation between these two 
approaches might involve ideas presented in \phmx .
 
Our noncritical M-theory provides a unified framework for the dynamics of 
two-dimensional noncritical strings.  Noncritical strings can be embedded into 
critical string theory via their relation to the topological strings on 
singular Calabi-Yau manifolds (see, \eg , \minaetal ).  Since the latter have 
been conjecturally related to a topological M-theory in seven dimensions 
\topom , it would be interesting to see whether an embedding of our 
noncritical M-theory into critical string/M-theory can shed light on the  
seven-dimensional topological M-theory.

Using the Fermi liquid picture, we have established that noncritical M-theory 
in $2+1$ dimensions can be defined.  However, many open questions 
clearly remain.  For example, it is unclear how to formulate this theory 
directly in terms of a matrix model.  An even more pressing challenge is 
to understand the effective spacetime description of noncritical M-theory 
vacua, in a language that directly refers to gravity in $2+1$ dimensions.  
Guided by noncritical string theory, we expect that such an effective 
spacetime gravity description indeed exists.  In this description, we expect 
a propagating degree of freedom -- the M-theory analog of the massless 
tachyon -- coupled to a gravitational sector.  In noncritical string theory, 
the relationship between the eigenvalue space and the spacetime Liouville 
dimension is known to be subtle, involving a nonlocal Laplace-like transform.  
Finding its analog in noncritical M-theory represents one of the main 
challenges.  It is also natural to ask what is the full spectrum of solitons 
in the theory, and in particular, whether or not the noncritical M-theory 
vacuum contains membranes.    

In its fermionic formulation, our noncritical M-theory is a rather unique 
theory, specified by the underlying infinite $\CW$ symmetry of its Lagrangian.%
\foot{This point emerged from discussions with Shamit Kachru.}
One natural extension, compatible with the $\CW$ symmetry, would be the 
addition of spin to the fermions.  Perhaps this possibility may be related to 
the existence of two different RR gauge fields in Type 0A theory.  In 
principle, one can also try to include the nonsinglet states,%
\foot{We thank Herman Verlinde for raising this issue.}   
although it is unclear -- in the absence of a direct matrix model formulation 
-- how they can be accommodated and what physical role they will play in the 
theory.%

As to the hope that noncritical M-theory may teach us valuable lessons about 
the mysterious aspects of M-theory, it is encouraging to see that this theory 
is described by an exactly solvable system.  Exact results for various 
physical observables are now in principle available, and the challenge is to 
interpret them and draw the corresponding lessons.  
The exact evaluation of the vacuum energy in M-theory (essentially, the 
cosmological constant of the vacuum) performed in this paper is an example, 
in which several surprising features have been observed.  

Our noncritical M-theory may also be of interest from the general viewpoint 
of quantum gravity.  In $2+1$ dimensions, there are essentially two successful 
approaches to quantum gravity, each with its own drawbacks.  The first one is 
the Chern-Simons formulation \refs{\ewcs,\carlip}, in which the topological 
nature of the theory is prominent.  However, it is difficult to include any 
propagating degrees of freedom in this framework.  The second possibility is 
to study compactifications of the full critical string/M theory to $2+1$ 
dimensions, for example on $AdS_3$.  This also defines a consistent quantum 
gravitational system, at the cost of carrying the entire baggage of the 
stringy and KK degrees of freedom.  The noncritical M-theory defined in this 
paper may represent a middle road to quantum gravity in $2+1$ dimensions, 
allowing a propagating degree of freedom but sharing some of the topological 
features with the Chern-Simons approach.  

It is worth pointing out that noncritical M-theory represents a framework in 
which the physical spacetime is an emergent property, available  only 
for those solutions of the underlying quantum mechanical system that admit 
a hydrodynamic description.  Moreover, this theory seems to be a realization 
of Mach's principle \mach :  The semiclassical physical spacetime is sustained 
by the collective motion of $N$ fundamental constituent fermions.  Without the 
constituents, there is no hydrodynamics of the Fermi liquid, and consequently 
no spacetime.  

\bigskip
\bigskip
\noindent{\bf Acknowledgements}
\medskip
We wish to thank Josh Friess, Eric Gimon, Shamit Kachru, Peter Shepard, 
and Herman Verlinde for useful discussions.  The results reported in this 
paper were presented at TASI on Particle Physics in Boulder in June 2005 
(by CAK), and at the Strings 2005 Conference in Toronto in July 2005 (by PH).  
We would like to thank the organizers of these meetings for their kind 
hospitality and the opportunity to present our results.  
This material is based upon work supported by NSF grant PHY-0244900, 
DOE grant DE-AC02-05CH11231, an NSF Graduate Research Fellowship, and 
the Berkeley Center for Theoretical Physics.  

\listrefs
\end